\documentclass[fleqn,usenatbib]{mnras}

\usepackage{CJKutf8}
\usepackage{orcidlink}

\usepackage{newtxtext,newtxmath}

\usepackage[T1]{fontenc}

\DeclareRobustCommand{\VAN}[3]{#2}
\let\VANthebibliography\thebibliography
\def\thebibliography{\DeclareRobustCommand{\VAN}[3]{##3}\VANthebibliography}

\usepackage{graphicx}	
\usepackage{amsmath}	
\usepackage{threeparttable}
\usepackage{academicons,xcolor}
\usepackage{caption}
\usepackage{ulem}

\newcommand{\revised}{}
\newcommand{\revisedtwo}{}
\newcommand{\revisedthree}{}

\newcommand{\eventname}{OGLE-2015-BLG-0845 }
\newcommand{\spitzer}{\textit{Spitzer} }

\newcommand{\DeltaChisrPX}{$\Delta \chi^{2} \simeq  340$}
\newcommand{\DeltaChisrTwoSPX}{$\Delta \chi^{2} \simeq  50$}

\newcommand{\SonePhot}{(V-I, I)_{\rm{S1}} = (1.46 \pm 0.05, 18.67 \pm 0.13)}
\newcommand{\StwoPhot}{(V-I, I)_{\rm{S2}} = (1.5\pm 0.4, 20.9\pm 0.5)}
\newcommand{\RCzero}{(V-I, I)_{\rm{RC,0}} = (1.06 \pm 0.06, 14.46 \pm 0.04)}
\newcommand{\RCfit}{(V-I, I)_{\rm{RC}} = (1.972\pm0.012, 15.82\pm 0.07)}
\newcommand{\Extinction}{(E_{\rm{S}}(V-I), A_{\rm{S}}(I)) = (0.91 \pm 0.06, 1.37 \pm 0.08)}
\newcommand{\SonePhotZero}{(V-I, M_{\rm{I}} )_{\rm{S1,0}} = (0.55 \pm 0.08, 2.7 \pm 0.3)}
\newcommand{\StwoPhotZero}{(V-I, M_{\rm{I}} )_{\rm{S2,0}} = (0.5 \pm 0.4, 5.0 \pm 0.6)}

\newcommand{\orcid}[1]{\href{https://orcid.org/#1}{\textcolor[HTML]{A6CE39}{\aiOrcid}}}

\title[Xallarap Event OB150845]{OGLE-2015-BLG-0845L: A low-mass M dwarf from the microlensing parallax and xallarap effects}

\author[Hu et al.]{
Zhecheng Hu (\begin{CJK*}{UTF8}{gbsn}胡哲程\end{CJK*})\orcidlink{0009-0000-6461-5256},$^{1}$\thanks{E-mail: hzc22@mails.tsinghua.edu.cn}
Wei Zhu (\begin{CJK*}{UTF8}{gbsn}祝伟\end{CJK*})\orcidlink{0000-0003-4027-4711},$^{1}$\thanks{E-mail:weizhu@tsinghua.edu.cn}\thanks{The Spitzer Team}
Andrew Gould,$^{2,3}$\footnotemark[3]
Andrzej Udalski\orcidlink{0000-0001-5207-5619},$^{4}$\thanks{The OGLE Collaboration}
\newauthor
Takahiro Sumi,$^{5}$\thanks{MOA Collaboration}
Ping Chen (\begin{CJK*}{UTF8}{gbsn}陈平\end{CJK*}),$^{6}$
Sebastiano Calchi Novati\orcidlink{0000-0002-7669-1069},$^{7}$\footnotemark[3]
Jennifer C. Yee\orcidlink{0000-0001-9481-7123},${^8}$\footnotemark[3]
\newauthor
Charles A. Beichman,$^{7}$\footnotemark[3]
Geoffery Bryden,$^{9}$\footnotemark[3]
Sean Carey\orcidlink{0000-0002-0221-6871},$^{7}$\footnotemark[3]
Michael Fausnaugh\orcidlink{0000-0002-9113-7162},$^{10}$\footnotemark[3]
\newauthor
B. Scott Gaudi\orcidlink{0000-0003-0395-9869},$^{3}$\footnotemark[3]
Calen B. Henderson\orcidlink{0000-0001-8877-9060},$^{7}$\footnotemark[3]
Yossi Shvartzvald\orcidlink{0000-0003-1525-5041},$^{6}$\footnotemark[3]
Benjamin Wibking\orcidlink{0000-0003-3175-2291},$^{11}$\footnotemark[3]
\newauthor
Przemek Mr\'oz\orcidlink{0000-0001-7016-1692},$^{4}$\footnotemark[4]
Jan Skowron\orcidlink{0000-0002-2335-1730},$^{4}$\footnotemark[4]
Rados{\l}aw Poleski\orcidlink{0000-0002-9245-6368},$^{4}$\footnotemark[4]
Micha{\l} K. Szyma\'nski\orcidlink{0000-0002-0548-8995},$^{4}$\footnotemark[4]
\newauthor
Igor Soszy\'nski\orcidlink{0000-0002-7777-0842},$^{4}$\footnotemark[4]
Pawe\l{} Pietrukowicz\orcidlink{0000-0002-2339-5899},$^{4}$\footnotemark[4]
Szymon Koz\l{}owski\orcidlink{0000-0003-4084-880X},$^{4}$\footnotemark[4]
Krzysztof Ulaczyk\orcidlink{0000-0001-6364-408X},$^{12}$\footnotemark[4]
\newauthor
Krzysztof A. Rybicki\orcidlink{0000-0002-9326-9329},$^{4,6}$\footnotemark[4]
Patryk Iwanek\orcidlink{0000-0002-6212-7221},$^{4}$\footnotemark[4]
Marcin Wrona\orcidlink{0000-0002-3051-274X},$^{4}$\footnotemark[4]
Mariusz Gromadzki\orcidlink{0000-0002-1650-1518},$^{4}$\footnotemark[4]
\newauthor
Fumio Abe,$^{13}$\footnotemark[5]
Richard Barry,$^{14}$\footnotemark[5]
David P. Bennett,$^{14,15}$\footnotemark[5]
Aparna Bhattacharya,$^{14,15}$\footnotemark[5]
Ian A. Bond,$^{16}$\footnotemark[5]
\newauthor
Hirosane Fujii,$^{13}$\footnotemark[5]
Akihiko Fukui,$^{17,18}$\footnotemark[5]
Ryusei Hamada,$^{5}$\footnotemark[5]
Yuki Hirao,$^{19}$\footnotemark[5]
Stela Ishitani Silva,$^{20,14}$\footnotemark[5]
\newauthor
Yoshitaka Itow,$^{13}$\footnotemark[5]
Rintaro Kirikawa,$^{5}$\footnotemark[5]
Naoki Koshimoto,$^{5}$\footnotemark[5]
Yutaka Matsubara,$^{13}$\footnotemark[5]
Shota Miyazaki,$^{21}$\footnotemark[5]
\newauthor
Yasushi Muraki,$^{13}$\footnotemark[5]
Greg Olmschenk,$^{14}$\footnotemark[5]
Cl\'ement Ranc,$^{22}$\footnotemark[5]
Nicholas J. Rattenbury,$^{23}$\footnotemark[5]
Yuki Satoh,$^{5}$\footnotemark[5]
\newauthor
Daisuke Suzuki,$^{5}$\footnotemark[5]
Mio Tomoyoshi,$^{5}$\footnotemark[5]
Paul. J. Tristram,$^{24}$\footnotemark[5]
Aikaterini Vandorou,$^{14,15}$\footnotemark[5]
Hibiki Yama,$^{5}$\footnotemark[5]
\newauthor
Kansuke Yamashita$^{5}$\footnotemark[5]
\newauthor
\\
Affiliations are listed at the end of the paper.
}

\date{Accepted XXX. Received YYY; in original form ZZZ}

\pubyear{2024}

\begin{document}
\label{firstpage}
\pagerange{\pageref{firstpage}--\pageref{lastpage}}
\maketitle

\begin{abstract}
We present the analysis of the microlensing event OGLE-2015-BLG-0845, which was affected by both the microlensing parallax and xallarap effects. The former was detected via the simultaneous observations from the ground and Spitzer, and the latter was caused by the orbital motion of the source star in a relatively close binary. The combination of these two effects led to a mass measurement of the lens object, revealing a low-mass ($0.14 \pm 0.05 M_{\sun}$) M-dwarf at the bulge distance ($7.6 \pm 1.0$ kpc). The source binary consists of a late F-type subgiant and a K-type dwarf of $\sim1.2 M_{\sun}$ and $\sim 0.9 M_{\sun}$, respectively, and the orbital period is $70 \pm 10$ days. OGLE-2015-BLG-0845 is the first single-lens event in which the lens mass is measured via the binarity of the source. Given the abundance of binary systems as potential microlensing sources, the xallarap effect may not be a rare phenomenon. Our work thus highlights the application of the xallarap effect in the mass determination of microlenses, and the same method can be used to identify isolated dark lenses.
\end{abstract}




\section{Introduction}
\label{sec:intro}

\begin{figure*}
    \includegraphics[width=1.8\columnwidth]{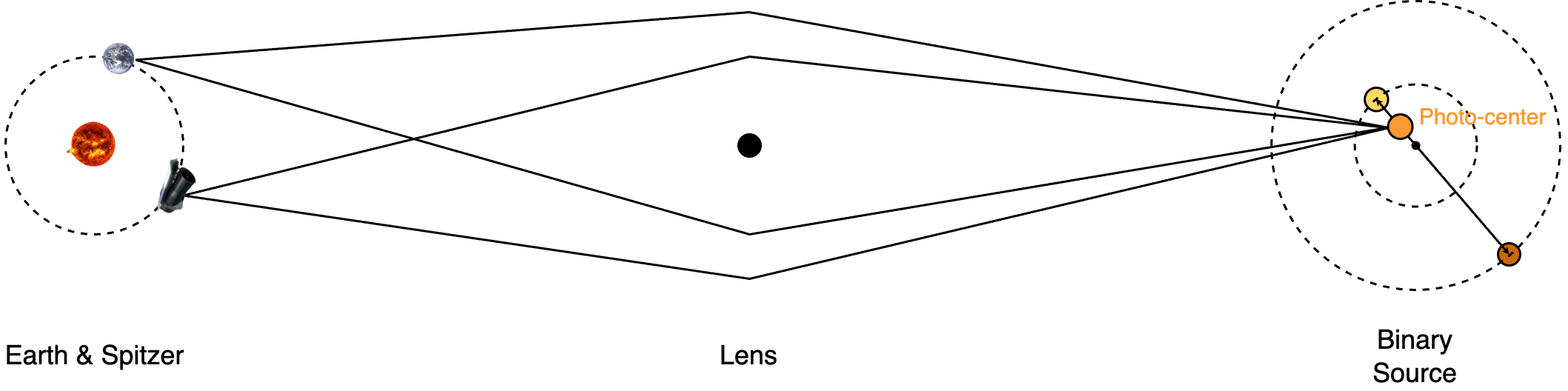}
    \caption{An illustration of the satellite parallax method and the xallarap effect in the microlensing phenomenon. Distances and physical sizes are not to scale. The microlensing source is a binary system, with the yellow circle indicating the primary star (i.e., primary source) and the brown circle indicating the secondary star (i.e., secondary source). The centre of light, namely the photocentre, of the source system is indicated by the orange circle. On the observer plane, Spitzer was in an Earth-trailing orbit and observed the same source simultaneously as the ground-based observatories.
    }
    \label{fig:xlrp_geo}
\end{figure*}

The microlensing effect \citep{Einstein1936_LensLike, Paczynski1986_Gravitational} is able to measure the mass of faint or even dark objects (or stellar systems) independent from their flux \citep[e.g.,][]{Gould1992_Extending}. So far, the microlensing technique has been used to determine the masses of dozens of isolated objects, including brown dwarfs \citep[e.g.,][]{Gould2009_Extreme, Zhu2016_Mass, Shvartzvald2019_Spitzer}, low-mass stars \citep[e.g.,][]{Chung2017_OGLE-2015-BLG-1482L, Zhu2017_Isolated, Shin2018_OGLE-2016-BLG-1045, Zang2020_Spitzer, Zang2020_SpitzerVLTI}, and stellar remnants \citep{Sahu2022_Isolated, Lam2022_isolated, Mroz2022_Systematic}, as well as another dozens of planetary systems \citep[e.g.][]{Gaudi2008, Bennett2015_Confirmation}.

Out of thousands of microlensing events that are discovered per year, only a small fraction of them allow one to determine the lens mass directly. This is because, in order to directly determine the lens mass, one needs to measure at least two out of three parameters, namely the angular Einstein radius, the microlensing parallax, and the lens flux \citep[e.g.,][]{Yee2015_Lens}. The angular Einstein radius measures the characteristic angular size of the microlensing phenomenon and is given by
\begin{equation}
    \theta_{\rm E} \equiv \sqrt{\kappa M_{\rm L} \pi_{\rm rel}} .
\end{equation}
Here $\kappa \equiv 4G/(c^2 \rm AU) \approx 8.14 \ {\rm mas}/M_\odot$ is a constant, $M_{\rm L}$ is the lens mass, and $\pi_{\rm rel} \equiv {\rm AU} (D_{\rm L}^{-1} - D_{\rm S}^{-1})$ is the lens--source relative parallax \citep{Gould2000_Natural}. \revised{The distances to the source and to the lens are $D_\mathrm{S}$ and $D_{\rm L}$, respectively.} The microlensing parallax measures the relative parallactic motion scaled to the angular Einstein radius \citep{Gould1992_Extending}
\begin{equation}
    \pi_{\rm E} \equiv \frac{\pi_{\rm rel}}{\theta_{\rm E}} .
\end{equation}
The lens flux measures the observed flux of the lens at the given distance. With a stellar mass--luminosity relation and a given extinction model, the lens flux measurement provides a relation between the lens mass and lens distance, similar to the other two quantities \citep[e.g.,][]{Bennett2015_Confirmation}. For faint or dark lenses, the lens flux is usually not detectable, and thus the only option towards a direct mass measurement is through the combination of angular Einstein radius and the microlensing parallax.

Although the microlensing parallax can be measured through the orbital motion of Earth around the Sun \citep{Gould1992_Extending}, such an annual parallax effect only applies to microlensing events with relatively long timescales \citep[e.g.,][]{Mao:2002, Poindexter2005_Systematic, Wyrzykowski2016_Black}.
The parameter $\pi_{\rm E}$ of an event could also be measured with at least two well-separated ($\sim {\rm AU}$) observatories\citep{Refsdal1966_Possibility, Gould1994_MACHO}. Between 2014 and 2019, \spitzer was used to observe more than five hundred microlensing events, and so far nearly a hundred of them with parallax solutions have been published \citep[e.g.,][]{CalchiNovati2015_Pathway, Zhu2017_Galactic}.

Several methods are available to measure the Einstein radius, but each has its own limitations. Observing the lens and source separately several years after the event could obtain the lens--source relative motion $\mu_{\rm rel}$, and then $\theta_{\rm E}$ is given by $\theta_{\rm E} = t_{\rm E} \mu_{\rm rel}$, where $t_{\rm E}$ is the Einstein crossing time \citep[e.g.][]{Alcock2001_Direct, Bennett2015_Confirmation, Gould2022_MASADA}. 
The angular Einstein radius can also be determined through the finite source effect, which appears when the source is close to or even crosses the caustic of the lens object/system \citep{Gould1994_Proper, Witt1994_Can, Nemiroff1994_Finite}, but in the case of single-lens events, the chance is low to have the finite source effect, except for extremely low-mass
lenses such as free-floating planets \citep[e.g.,][]{Mroz2018_Neptune, Gould2022_Free}. The astrometric microlensing has also been proposed \citep{Hog1995_MACHO}, and recently realized \citep{Sahu2022_Isolated, Lam2022_isolated}, as a method to directly measure $\theta_{\rm E}$ during the course of the microlensing event. Additionally, interferometric observations may be able to directly resolve the multiple microlensing images and thus determine $\theta_{\rm E}$ \citep{Delplancke2001_Resolving, Dong:2019, Cassan:2022}. These latter two methods have not been applied widely because they require high precisions and/or lucky observing conditions. 

When the source is in a binary system, its orbital motion around the centre of mass may lead to changes in the relative positions among the microlensing objects, which also revises the light curve \citep{Griest1992_Effect}. This so-called microlensing xallarap effect can also be used to measure the angular Einstein radius \citep{Han1997_Einstein}.
Specifically, the xallarap parameter, $\xi_\mathrm{E}$, is related to $\theta_\mathrm{E}$ via
\begin{equation}
\label{eq:xlrp_def}
    \xi_{\mathrm{E}} \equiv \frac{a_{\mathrm{S}}}{D_{\mathrm{S}} \theta_{\mathrm{E}}}=\frac{a_{\mathrm{S}}}{\hat{r}_{\mathrm{E}}} .
\end{equation}
Here $a_\mathrm{S}$ is the semi-major axis of the motion of the source star (or the photocentre of the source binary) around the barycentre of the source binary system, and $\hat{r}_\mathrm{E}$ is the projected Einstein radius in the source plane.
Once $\xi_\mathrm{E}$ and $a_{\rm S}$ are measured, the xallarap effect provides another way to measure $\theta_{\rm E}$. 

Given the abundance of stellar binaries \citep[e.g.,][]{Duchene2013_Stellar}, a substantial fraction of events might have been affected by the xallarap effect. \citet{Poindexter2005_Systematic} searched for the xallarap effect in 22 microlensing events with relatively long ($t_{\rm E} \gtrsim 70$ d) time scales and found that about 23\% of them might have been strongly affected by the xallarap effect, although the fraction may be reduced once the shorter but more abundant events are taken into account. The impact of xallarap effect has also been regularly investigated in microlensing events that contain potentially planetary signals \citep[e.g.,][]{Furusawa2013_MOA-2010-BLG-328Lb, Miyazaki2020_OGLE-2013-BLG-0911Lb, Rota2021_MOA-2006-BLG-074, Satoh2023_OGLE-2019-BLG-0825, Yang:2024, Ryu2024_Systematic}.

In this work, we present the analysis of the microlensing event \eventname. This event shows both xallarap and parallax effects, as illustrated in Figure~\ref{fig:xlrp_geo}, which allow us to determine directly the lens mass and distance. The observations of \eventname are presented in Section~\ref{sec:obs}, the detailed modelings of the event are given in Section~\ref{sec:model}, and the physical properties of the lens and source are given in Section~\ref{sec:phy_interp}. A brief discussion of the results is given in Section~\ref{sec:discuss}.

\section{Observations}
\label{sec:obs}

The microlensing event \eventname was first identified by the OGLE-IV collaboration on UT 15:51, 28 April 2015. With equatorial coordinates $\rm{(RA, Dec)}_{2000} = (18^{\rm{h}}04^{\rm{m}}21\fs 29, -31\degr 34\arcmin 50\farcs 0)$ and Galactic coordinates $(l, b)_{2000} = (-0\fdg 25, -4\fdg 82)$, this event was located inside the field BLG514 of OGLE-IV survey and received three observations per day from the 1.3\,m Warsaw Telescope at the Las Campanas Observatory in Chile \citep{Udalski2003_Optical,Udalski2015_OGLE-IV}. The OGLE observations were mostly taken in the $I$ band, but $V$ band observations were also taken at the cadence of a few days in order to provide colour information of the microlensing source. The OGLE data were reduced using the software developed by \citet{Wozniak2000_Difference} and \citet{Udalski2003_Optical}, which was based on the Difference Image Analysis (DIA) technique of \citet{Alard1998_Method}.

This event was observed by the Spitzer Space Telescope as part of the 2015 microlensing campaign \citep{Udalski2015_SPITZERAS,Yee2015_Criteria,Zhu2017_Galactic}. It was selected for Spitzer observations on June 8, 2015, as a ``secret event'' and then announced as a Spitzer target on June 12, 2015. We refer to \citet{Yee2015_Criteria} for the details of the observing protocol of the Spitzer microlensing program. The Spitzer observations started on the same date and stopped on July 19, 2015, when the event moved out of the visibility window of Spitzer. There were in total 148 observations taken by Spitzer. All Spitzer observations were reduced by the customized software specifically for the microlensing program \citet{CalchiNovati2015_Spitzer}.

Event \eventname was also observed by the Microlensing Observations in Astrophysics \citep[MOA,][]{Bond2001_Real-time,Sako2008_MOA-cam3}. It is called MOA-2015-BLG-277 in the MOA database. Follow-up observations from various small telescopes were also taken for the purpose of detecting planetary signals. Because of the unknown systematics and/or the relatively short time baseline, the MOA data and the follow-up data are not included in the modeling of the \revised{subtle signals such as the} parallax and xallarap signals, \revised{as is commonly done in the modeling of other microlensing events \citep[e.g.,][]{Mroz2022_Systematic}. In the case of \eventname, because there are about twice as many observations from MOA as from OGLE, the more informative OGLE data would be substantially downweighted if one were to include the MOA data in the modeling. We have nevertheless included the MOA data in the single source binary lens (2L1S) and binary source without xallarap effect (1L2L static) modelings, given that the potential signals from such models are relatively short-timescale (see Appendix~\ref{app:a1_2L1S} and Appendix~\ref{app:a2_static_1L2S}).} 

\section{Modeling}
\label{sec:model}

The light curve of \eventname was first modeled using the standard \citet{Paczynski1986_Gravitational} curve
\begin{equation}
f(t) = f_{\mathrm{s}} A(u[t]) + f_{\mathrm{b}}
\end{equation}
with the single-lens point-source, or 1L1S
\footnote{This notation follows the convention that is commonly used in the microlensing literature. The numbers in front of ``L'' and ``S'' indicates the numbers of objects in the lens and the source systems, respectively, that directly participate in the microlensing phenomenon in a detectable way. In other words, with ``1L'' (or ``1S'') it does not necessarily mean that the source (or lens) object is physically single.}
, microlensing magnification given by
\begin{align}
A (u) = \frac{u^{2}+2}{u \sqrt{u^{2} + 4}} .
\end{align}
Here $f_{\rm s}$ and $f_{\rm b}$, are the flux of the source star and the flux of the blending object, respectively. The quantity, $u(t)$, is the separation normalized to $\theta_\mathrm{E}$ between the lens and the source at a given epoch $t$. For the standard \citet{Paczynski1986_Gravitational} curve,
\begin{equation}
u(t)=\sqrt{\tau^{2}+\beta^{2}} ,
\end{equation}
with
\begin{equation} \label{eqn:rectilinear}
\tau(t) \equiv \frac{t-t_0}{t_{\mathrm{E}}}, \quad \beta(t) \equiv u_0 .
\end{equation}
Here $t_0$ is the time when the projected separation between the lens and the source reaches the minimum, $u_0$ is this minimum separation, and $t_{\rm E}$ is the Einstein crossing time.

\begin{figure}

	\includegraphics[width=0.95\columnwidth]{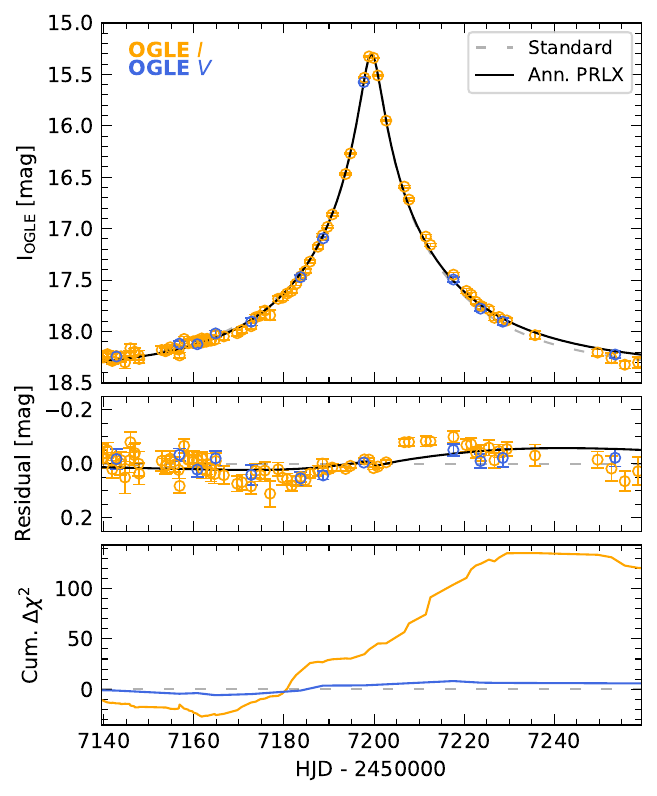}
    \caption{
    OGLE-IV data of \eventname and the best-fit 1L1S standard and annual parallax models. The top panel displays the OGLE $I$- and $V$-band data in black and yellow points, respectively. The best-fit standard model and the parallax model are denoted by the grey dashed and the black solid curves, respectively. The middle panel illustrates the residuals of the data and the parallax model with respect to the standard model. The bottom panel shows the cumulative $\Delta \chi^2 \equiv \chi^2_{\rm STD} - \chi^2_{\rm Anu. PRLX}$ between the standard and parallax models as a function of time. Although the best-fit parallax model improved the model $\chi^2$ by $>100$ over the standard 1L1S model, there remain long-term variations in the residuals that cannot be well fit.
    }
    \label{fig:lc_std_prlx}
\end{figure}

The standard \citet{Paczynski1986_Gravitational} curve cannot fit the light curve of \eventname very well, as shown in Figure~\ref{fig:lc_std_prlx}. While these deviations could be modeled by the annual parallax effect reasonably well, it was soon recognized that the parallax parameters from the annual parallax effect did not match those from the satellite parallax effect (Section~\ref{sec:tension}). This led to the inclusion of additional effects, including the xallarap effect (Section~\ref{subsec:xlrp_model}). The binary lens, as well as the static binary source model, was also considered, although it could not fit the data as well as the xallarap model (see Appendix \ref{app:a1_2L1S} and Appendix~\ref{app:a2_static_1L2S}).

In all the modelings, we have made use of the \texttt{emcee} package \citep{Foreman-Mackey2013_emcee} to execute a Markov Chain Monte Carlo (MCMC) analysis. The number of walkers and burn-in steps have been adjusted so as to make sure the chain has converged.

\revisedtwo{The color constraint, which has been frequently used to infer the source flux in Spitzer band based on the known source flux values in the ground-based bandpasses \citep[e.g.,][]{CalchiNovati2015_Spitzer, Yee2015_Criteria}, is not used in the present event. The color--color relation is usually constructed based on field stars of similar colors, so its applicability to binary stars is questionable. One may in principle derive two separate color--color relations for the two components of the source binary, but this is difficult to implement as the source binary is not well constrained by the ground-based data alone (see Section~\ref{sec:1L2S}). Instead of such statistical test as the color constraint, we will use more direct tests to verify the validity of the Spitzer data.}

\subsection{The Tension Between 1L1S Annual and Space-based Parallax Model} \label{sec:tension}

The accelerated motion of the Earth can lead to a distortion on the light curve, known as the annual parallax effect \citep[e.g.,][]{Gould1992_Extending, Alcock1995_First}. This distortion adds displacement to the rectilinear motion (Equation~\ref{eqn:rectilinear}) of the \citet{Paczynski1986_Gravitational} model
\begin{equation}
(\delta \tau_\mathrm{p}, \delta \beta_\mathrm{p})=\left(\boldsymbol{\pi}_{\mathrm{E}} \cdot \Delta \boldsymbol{s}, \boldsymbol{\pi}_{\mathrm{E}} \times \Delta \boldsymbol{s}\right).
\end{equation}
Here $\Delta \boldsymbol{s}$ is the offset of the Earth-to-Sun vector in AU \citep{Gould2004_Resolution}. The microlensing parallax vector is defined as
\begin{equation}
\boldsymbol{\pi}_{\mathrm{E}}=\frac{\pi_{\mathrm{rel}}}{\theta_{\mathrm{E}}} \frac{\boldsymbol{\mu}_{\mathrm{rel}}}{\mu_{\mathrm{rel}}} .
\end{equation}
Here $\vec{\mu}_{\rm{rel}}$ is the relative proper motion between the lens and source, respectively \citep{Gould2000_Natural}. 

The microlensing parallax can also be measured through simultaneous observations of the same event from at least two well--separated observatories/satellites \citep{Refsdal1966_Possibility, Gould1994_MACHO, Zhu2017_Isolated}.
The two observatories---namely Spitzer and Earth in the present case---obtain light curves that are different in both the event peak time $t_0$ and the impact parameter $u_0$. Making use of the known projected separation between the two observatories on the celestial plane, $D_\perp$, the parallax vector $\boldsymbol{\pi}_\mathrm{E}$ can be estimated as
\begin{equation}     
\boldsymbol{\pi}_{\mathrm{E}} \approx \frac{\mathrm{AU}}{D_{\perp}}(\tau_{\mathrm{sat}}-\tau_\oplus, \beta_{\mathrm{sat}}-\beta_\oplus) .
\end{equation} 
Here the subscripts ``$\mathrm{sat}$'' and ``$\oplus$'' represent the quantities of the satellite and Earth, respectively. 

\begin{figure}
\includegraphics[width=1.\columnwidth]{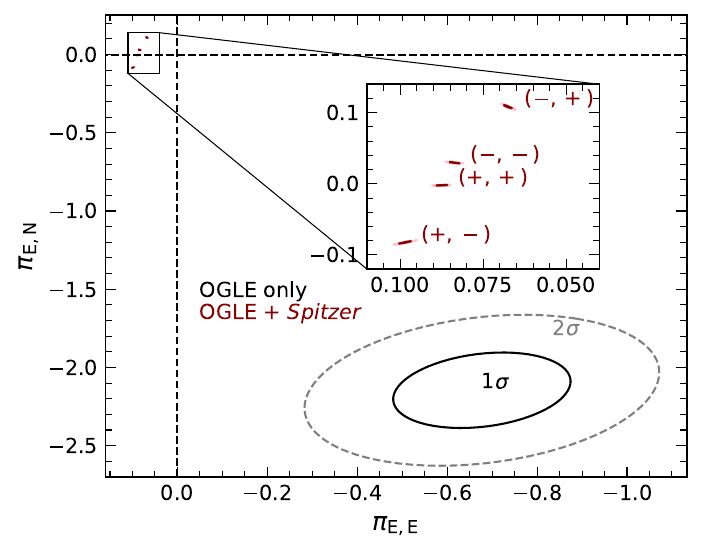}
\caption{
The tension between $\pi_\mathrm{E}$ as derived by fitting OGLE-only data and joint fitting OGLE and \spitzer data of the microlensing event \eventname.
The 1-sigma regions are shown as dark solid ellipses, while the 2-sigma regions are shown as lighter dashed ellipses.
For space-based parallax, all four degenerate solutions are shown while for annual parallax, only the $u_{0}+$ solution is shown.
The $u_0-$ solution for the OGLE-only fitting is not shown because the $\pi_{\rm{E}}$ value is close to the positive solutions, which does not change the tension.
}
\label{fig:piE-tension}
\end{figure}

The two approaches should yield consistent parallax parameters, provided that the deviation from the standard \citet{Paczynski1986_Gravitational} model is indeed due to the annual parallax effect. As shown in Figure~\ref{fig:piE-tension}, for \eventname, the constraints on microlensing parallax from fitting OGLE data alone significantly differ from the constraints from the joint fitting of OGLE and Spitzer data, no matter which of the four degenerate solutions is considered. 
Therefore, other explanations are required to explain the deviation in the light curve from the standard \citet{Paczynski1986_Gravitational} model.

\begin{figure*}
\includegraphics[width=\textwidth]{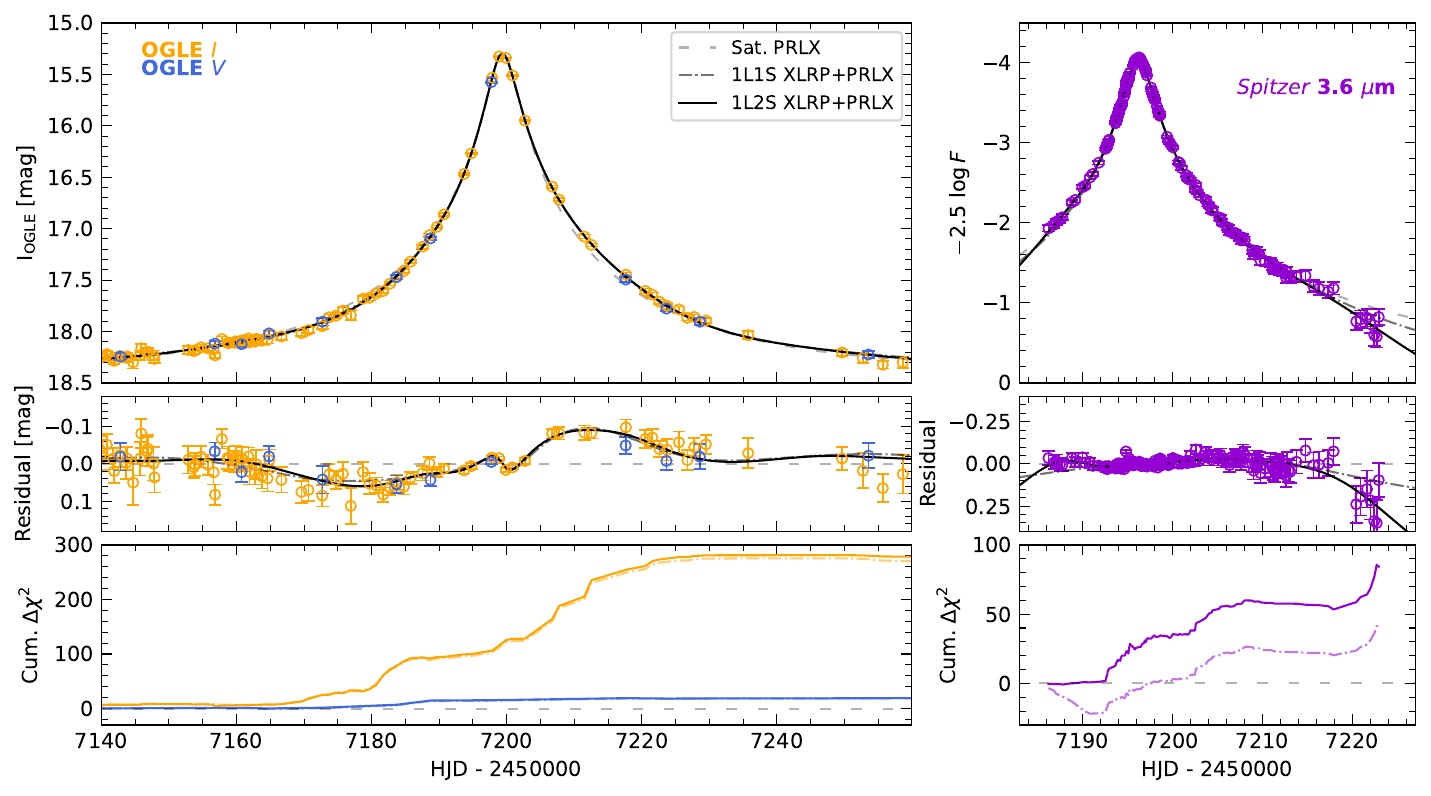}
\caption{
OGLE $I$ and $V$ band light curves (left panels) and \spitzer 3.6\,$\mu$m light curve (right panels) of the microlensing event \eventname. In each panel, the best-fit 1L1S parallax model, 1L1S parallax + xallarap model, and the 1L2S parallax + xallarap model are shown as dashed gray line, dash-dotted line, and solid line, respectively. The OGLE $V$-band data have been aligned to the OGLE $I$-band data based on the best-fit flux parameters, and the \spitzer flux values are shown in logarithmic scale (quasi-magnitudes).
The top panels show the observed data and the best-fit models, and the middle panels show the residuals of data and the other models relative to the 1L1S parallax model. In the lower panels, we show the cumulative $\Delta \chi^2$ distribution of the other two models relative to the 1L1S parallax model. 
}
\label{fig:lc-all}
\end{figure*}

\subsection{1L1S Xallarap + Parallax Model}
\label{subsec:xlrp_model}

The xallarap effect is the reflection of the orbital motion of the source \revised{star/system} on the light curve, as is shown in Figure~\ref{fig:xlrp_geo}.
\revised{Whenever the xallarap effect is invoked, it almost always assumes that the microlensing source is a binary (or higher multiple). By including only one bright source in the xallarap modeling, we have assumed that the source binary is unresolved or that the secondary component is very faint/dark, which will be checked at the end of this subsection.
\footnote{In microlensing, a binary source being unresolved means that $\xi_{\rm E} \ll \theta_{\rm E}$ and $\xi_{\rm E} \ll u_0$.}
In either case, the ``source'' here refers to the photocenter of the binary system rather than any single star.}
For simplicity, the orbit of the source is assumed to be circular, and the impact of an elliptical orbit is discussed in the section \ref{subsec:ecc}.

The xallarap effect introduces five new parameters, $(P_\xi, A_\xi, B_\xi, F_\xi, G_\xi)$. Here $P_\xi$
is the orbital period of the source binary, and the other four parameters are variants of the classical Thiele-Innes elements that are widely used in astrometry
\begin{equation} \label{eqn:thiele-innes}
\left\{\begin{array}{l}
A_\xi=\xi_\mathrm{E} (\cos \phi_\xi \cos (-\theta_\xi)-\sin \phi_\xi \sin (-\theta_\xi) \cos i_\xi) \\
B_\xi=\xi_\mathrm{E} (\cos \phi_\xi \sin (-\theta_\xi)+\sin \phi_\xi \cos (-\theta_\xi) \cos i_\xi) \\
F_\xi=\xi_\mathrm{E} (-\sin \phi_\xi \cos (-\theta_\xi)-\cos \phi_\xi \sin (-\theta_\xi) \cos i_\xi) \\
G_\xi=\xi_\mathrm{E} (-\sin \phi_\xi \sin (-\theta_\xi)+\cos \phi_\xi \cos (-\theta_\xi) \cos i_\xi)
\end{array}\right.
\end{equation}
Here $i_\xi$ is the orbital inclination, $\phi_\xi$ is the phase of the source relative to the ascending node at the reference epoch $t_{\rm ref}$, and $\theta_\xi$ measures the ascending node relative to the lens--source relative trajectory projected onto the source plane. The amplitude of the xallarap motion, $\xi_{\rm E}$, is the semi-major axis of the binary motion scaled to the projected Einstein ring radius, as is defined in Equation \ref{eq:xlrp_def}.

The xallarap effect leads to displacement in the lens-source relative trajectory by the amounts
\begin{equation}
\label{eq:projection}
\left(\begin{array}{c}
\delta \tau_\mathrm{x} \\
\delta \beta_\mathrm{x}
\end{array}\right)=\left(\begin{array}{cc}
A_\xi & F_\xi \\
B_\xi & G_\xi
\end{array}\right) \cdot \Delta \boldsymbol{S}
\end{equation}
Following the convention in the modeling of the parallax effect \citep{Gould2004_Resolution}, we have introduced $\Delta \boldsymbol{S}$ to measure the displacement between the actual position of the source under binary motion and an imaginary source under linear motion
\begin{equation} \label{eqn:primary-source}
\Delta \boldsymbol{S} = (t-t_{\rm ref}) \boldsymbol{v_{\rm ref}} - [\boldsymbol{S}(t) - \boldsymbol{S}(t_{\rm ref}) ] .
\end{equation}
The binary motion of the source, $\boldsymbol{S}(t)$, is given by the circular motion starting from a zero phase with an amplitude of unity, and the velocity vector, $\boldsymbol{v_{\rm ref}}$, is the instantaneous velocity of the source \revised{relative to the center of mass of the source binary} at reference epoch $t_{\rm ref}$.

The distorted lens--source relative trajectory in the presence of both parallax and xallarap effects can be described as
\begin{equation}
\left\{\begin{array}{l}
\tau = \tau_{\rm{std}} + \delta \tau_{\rm{p}} + \delta \tau_{\rm{x}} \\
\beta = \beta_{\rm{std}} + \delta \beta_{\rm{p}} + \delta \beta_{\rm{x}}
\end{array}
\right. .
\end{equation}
where $\tau_{\rm{std}}$ and $\beta_{\rm{std}}$ describe the trajectory in the standard \citet{Paczynski1986_Gravitational} model.

The best-fit 1L1S xallarap + parallax model is shown in Figure~\ref{fig:lc-all}, and the best-fit parameters and the associated uncertainties are given in Table~\ref{tab:best-fit}. The amplitude of the xallarap parameter, $\xi_{\rm E}$, derived from the xallarap model, is also provided in the same table. Note that in this derivation we have ignored the nonuniform prior introduced by the Thiele--Innes parameterization of the xallarap model, and we have confirmed that the values would only be marginally revised if a nonuniform prior were imposed.

The 1L1S Xallarap + Parallax model significantly better describes the light curve by \DeltaChisrPX, with respect to the 1L1S Parallax model and removes the tension in parallax constraints between annual and space-based parallax-only models.
However, from Table~\ref{tab:best-fit}, we can see that the $\xi_\mathrm{E}$ is comparable to the amplitude of $u_{0}$.
The secondary source, which is fainter and presumably less massive, is expected to experience a binary motion with a larger amplitude and thus produce non-negligible features in the light curve. Therefore, it is reasonable to introduce the secondary source as a luminous component in the microlensing event.

\begin{table*}
\begin{threeparttable}
\rotatebox{-90}{
\begin{minipage}{\textheight}
\caption{Parameter values and uncertainties of the best-fit 1L1S and 1L2S xallarap + parallax solutions for \eventname. }
\renewcommand{\arraystretch}{1.4}
\begin{tabular}{lcccc|cccc}
\hline\hline
                        & \multicolumn{4}{c|}{1L1S XLRP + PRLX}                                                     & \multicolumn{4}{c}{1L2S XLRP + PRLX}                                                      \\ \hline
Solution Type           & $(+,+)$              & $(+,-)$              & $(-,+)$              & $(-,-)$              & $(+,+)$              & $(+,-)$              & $(-,+)$              & $(-,-)$              \\ \hline
$t_0$ $^a$    & $7199.385 \pm 0.014$ & $7199.381 \pm 0.015$ & $7199.382 \pm 0.016$ & $7199.388 \pm 0.014$ & $7199.293 \pm 0.018$ & $7199.278 \pm 0.017$ & $7199.427 \pm 0.014$ & $7199.432 \pm 0.013$ \\
$u_0$                   & $0.05 \pm 0.003$     & $0.0436 \pm 0.0025$  & $-0.044 \pm 0.003$   & $-0.05 \pm 0.003$    & $0.049 \pm 0.006$    & $0.046 \pm 0.004$    & $-0.043 \pm 0.004$   & $-0.047 \pm 0.004$   \\
$t_\mathrm{E}$ (days)   & $42.1 \pm 2.2$       & $47.9 \pm 2.4$       & $47.0 \pm 2.5$       & $42 \pm 2$           & $46 \pm 5$           & $49 \pm 3$           & $46 \pm 4$           & $43 \pm 4$           \\
$\pi_\mathrm{E,N}$      & $0.0 \pm 0.0013$     & $-0.063 \pm 0.004$   & $0.084 \pm 0.005$    & $0.0231 \pm 0.0018$  & $-0.0085 \pm 0.0013$ & $-0.074 \pm 0.007$   & $0.08 \pm 0.007$     & $0.0125 \pm 0.0015$  \\
$\pi_\mathrm{E,E}$      & $0.077 \pm 0.004$    & $0.077 \pm 0.004$    & $0.054 \pm 0.004$    & $0.074 \pm 0.004$    & $0.078 \pm 0.008$    & $0.071 \pm 0.006$    & $0.065 \pm 0.006$    & $0.07 \pm 0.006$     \\
$p_\xi$ (days)          & $41 \pm 4$           & $46 \pm 4$           & $44 \pm 5$           & $41 \pm 5$           & $74 \pm 8$           & $76 \pm 7$           & $66 \pm 9$           & $63 \pm 9$           \\
$A_\xi$ ($10^{-2}$)     & $-0.01 \pm 0.003$    & $-0.013 \pm 0.004$   & $-0.014 \pm 0.005$   & $-0.011 \pm 0.003$   & $-0.065 \pm 0.017$   & $-0.069 \pm 0.016$   & $-0.044 \pm 0.012$   & $-0.041 \pm 0.012$   \\
$B_\xi$ ($10^{-2}$)     & $-0.011 \pm 0.008$   & $-0.018 \pm 0.016$   & $0.019 \pm 0.018$    & $0.018 \pm 0.011$    & $0.16 \pm 0.04$      & $0.16 \pm 0.03$      & $0.15 \pm 0.04$      & $0.14 \pm 0.04$      \\
$F_\xi$ ($10^{-2}$)     & $-0.005 \pm 0.005$   & $0.003 \pm 0.004$    & $-0.001 \pm 0.004$   & $-0.007 \pm 0.005$   & $0.029 \pm 0.019$    & $0.026 \pm 0.017$    & $0.02 \pm 0.015$     & $0.01 \pm 0.016$     \\
$G_\xi$ ($10^{-2}$)     & $-0.061 \pm 0.019$   & $-0.062 \pm 0.007$   & $0.065 \pm 0.008$    & $0.063 \pm 0.010$     & $0.07 \pm 0.03$      & $0.09 \pm 0.03$      & $0.057 \pm 0.025$    & $0.06 \pm 0.03$      \\
\hline
$q_\xi$                 &            ---       & ---                  & ---                   & ---                  & $0.33 \pm 0.16$      & $0.28 \pm 0.13$      & $0.3 \pm 0.2$        & $0.26 \pm 0.19$      \\
$q_\mathrm{f,OGLE}$     &            ---       & ---                  & ---                   & ---                   & $0.14 \pm 0.07$      & $0.11 \pm 0.07$      & $0.18 \pm 0.08$      & $0.17 \pm 0.05$      \\
$q_\mathrm{f,OGLE,V}$   &            ---       & ---                  & ---                   & ---                   & $0.15 \pm 0.08$      & $0.14 \pm 0.08$      & $0.18 \pm 0.08$      & $0.17 \pm 0.06$      \\
$q_\mathrm{f,Spitzer}$  &            ---       & ---                  & ---                   & ---                   & $1.1 \pm 0.5$        & $1.2 \pm 0.3$        & $0.5 \pm 0.3$        & $0.6 \pm 0.2$        \\
$\phi_\xi^{b}$ (deg)       & $-80 \pm 110$        & $-74 \pm 25$         & $107 \pm 12$         & $107 \pm 13$         & $160 \pm 6$          & $156 \pm 4$          & $162 \pm 5$          & $160 \pm 5$          \\
$\theta_\xi^{b}$ (deg)     & $97 \pm 5$           & $91 \pm 4$           & $86 \pm 4$           & $81 \pm 5$           & $73.8 \pm 2.5$       & $74.3 \pm 2.0$       & $77.6 \pm 1.4$       & $77.2 \pm 1.7$       \\
$i_\xi^{b}$ (deg)          & $82 \pm 5$           & $77.7 \pm 2.5$       & $101 \pm 3$          & $97 \pm 4$           & $106 \pm 4$          & $106 \pm 3$          & $102 \pm 4$          & $99 \pm 4$           \\
$\xi_\mathrm{E}^{b}$  & $0.063 \pm 0.009$    & $0.064 \pm 0.011$    & $0.068 \pm 0.014$    & $0.066 \pm 0.012$    & $0.18 \pm 0.05$      & $0.19 \pm 0.04$      & $0.17 \pm 0.05$      & $0.16 \pm 0.05$      \\
Blend fraction                   & $0.19 \pm 0.05$      & $0.24 \pm 0.04$      & $0.21 \pm 0.05$      & $0.13 \pm 0.05$      & $0.03 \pm 0.14$      & $0.03 \pm 0.12$      & $0.11 \pm 0.10$       & $0.13 \pm 0.09$      \\
$I_{\rm{s}}$                     & $18.61 \pm 0.06$     & $18.69 \pm 0.06$     & $18.64 \pm 0.06$     & $18.53 \pm 0.06$     & $18.42 \pm 0.16$     & $18.38 \pm 0.13$     & $18.75 \pm 0.15$     & $18.55 \pm 0.12$     \\
$\chi^2$/dof            & 1219.38/1126         & 1193.15/1126         & 1191.66/1126         & 1219.89/1126        & 1148.25/1122         & 1145.36/1122         & 1148.09/1122         & 1143.39/1122         \\
\hline
\end{tabular}
\begin{tablenotes}
    \item{NOTE. \\
    $^a$ Defined in HJD-2450000. In all fittings, we set the reference time to $t_{0, {\rm par}} = t_{\rm ref} = 2457200$.

    $^{b}$ These are derived based on Equations~(\ref{eqn:thiele-innes}). 
    }
\end{tablenotes}
\label{tab:best-fit}
\end{minipage}
}
\end{threeparttable}
\end{table*}

\subsection{1L2S Xallarap + Parallax Model} \label{sec:1L2S}

In order to add a luminous secondary source, at least two new parameters are introduced into the modeling: the mass ratio and the wavelength-dependent flux ratio of the secondary source to the primary source, denoted as $q_{\xi}$ and $q_{\rm{f,}\lambda}$, respectively. Note that the flux ratio parameter is different for different filters. For this modeling, we have included the $I$-band and $V$-band data of the OGLE-IV survey and the Spitzer data, so three flux ratio parameters are included.

The microlensing magnification from this 1L2S model is given by 
\begin{equation}
A_{2\rm{s, } \lambda}(u_1, u_2) = \frac{A(u_{1}) + q_{\rm{f,} \lambda} A(u_{2})}{1 + q_{\rm{f, } \lambda}} .
\end{equation}
Here $u_1$ ($u_2$) is the projected separation between the lens and the primary (secondary) source. The position of the primary source is determined in the same way as in Section~\ref{subsec:xlrp_model}. For the secondary source, its positional offset relative to the rectilinear motion is given by 
\begin{equation}
\begin{aligned}
\Delta \boldsymbol{S}_{2} = (t- t_{\rm{ref}}) \boldsymbol{v}_{\rm{ref, 1}} + \boldsymbol{S}_{1}(t_{\rm ref}) - \frac{1}{q_{\xi}} \boldsymbol{S}_{1}(t) \\
\end{aligned}
\end{equation}
Here the subscripts 1 and 2 denote the primary and secondary source, respectively. Similar to that for the primary source (Equation~\ref{eqn:primary-source}), the first two terms transform the trajectory from the barycentric frame to the frame centred on the primary source. The last term calculates the orbit of the secondary source, which is $1 / q_{\xi}$ times larger than that of the primary source and on the opposite side of the barycentre. The parallax effect is included in the same way as in the 1L1S case.

As shown in Figure~\ref{fig:lc-all} and further detailed in Table~\ref{tab:best-fit}, the best-fit 1L2S Xallarap+Parallax model provides a better match to the data by \DeltaChisrTwoSPX, with \spitzer data contributing the majority of this improvement. Although this makes the 1L2S model suspicious at first sight, particularly given that some fraction of the \spitzer data is known to contain systematics \citep[e.g.,][]{Zhu2017_Galactic, Koshimoto:2020}, there are several pieces of evidence suggesting that the best-fit 1L2S solution is plausible. First, the 1L2S solution is inevitable according to the best-fit 1L1S solution, as has been explained in Section~\ref{subsec:xlrp_model}. Second, the redder secondary source, as inferred from the binary flux ratios in different bandpasses, is consistent with the estimated binary mass ratio directly inferred from the light curve fitting, where more details are described in \ref{subsec:source_prop}, so this binary solution is astrophysically plausible. 

\begin{figure}
\includegraphics[width=1.\columnwidth]{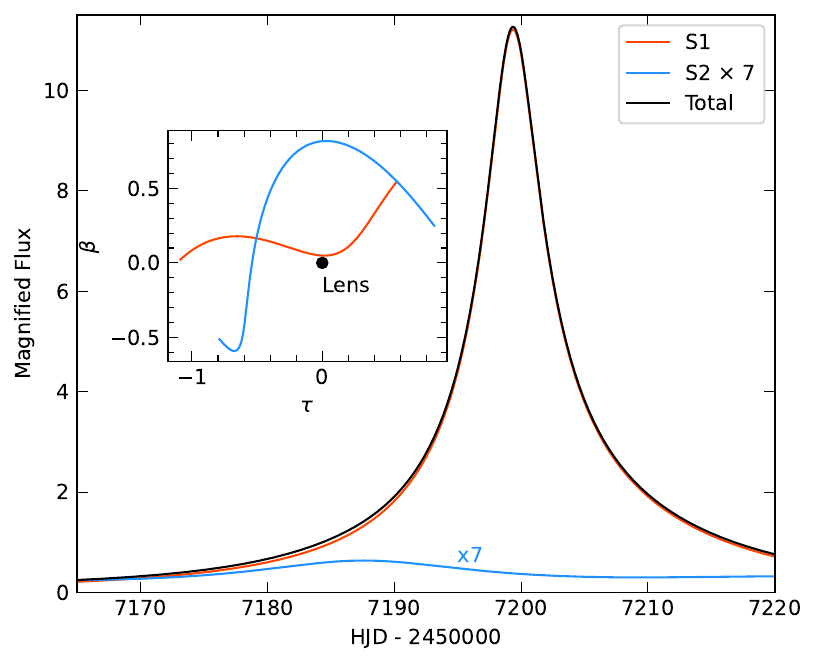}
\caption{
The flux variations of the two source components, S1 and S2, in the 1L2S model. The red and blue curves show the pure flux curve of the S1 and S2, respectively, and the black curve shows the combined result. For illustration purposes, only the $(+,+)$ solution is shown, and the flux variations of S2 have been amplified by a factor of seven. The inset shows the trajectories of the two sources with respect to the lens object. The two sources move from the left to the right.
}
\label{fig:magnification}
\end{figure}

Some of the derived parameters in the 1L2S solution appear different from those in the 1L1S solution. In particular, the orbital period of the source binary, $P_{\xi}$, changes from $\sim40\,$d to 60--80\,d, and the xallarap parameter, $\xi_{\rm{E}}$, changes from $\sim0.06$ to $\sim 0.2$. The change in $\xi_{\rm{E}}$ is due to the different definitions of this parameter in the two models. In the 1L1S model, $\xi_{\rm{E}}$ is defined by the motion of the centre of light (i.e., photocentre) relative to the centre of mass of the binary, whereas in the 1L2S model the same parameter refers to the motion of the primary source relative to the centre of mass of the binary. The photocentre is closer to the barycentre than the primary source by a factor of $(q_\xi-q_{\rm f})/(q_\xi+q_\xi q_{\rm f})$. Using the values in Table~\ref{tab:best-fit}, we find this factor to be $\sim 0.4$. Together with the statistical uncertainties, this can basically explain the change of $\xi_{\rm E}$. The change in the binary orbit period is related to the morphology of the xallarap signal. As shown in the left panel of Figure~\ref{fig:lc-all}, the xallarap signal appears to be wave-like once the standard 1L1S model is subtracted. This wave-like feature can be approximated by the summation of two lower-frequency (thus longer period) terms, as illustrated in Figure~\ref{fig:magnification}. That figure shows the flux variations in the lensing process while the contributions from the two source stars are shown separately. Here the flux variations are given by
\begin{equation}
\Delta F_1 = \frac{f_s}{1+q_f} (A_1-1),\quad \Delta F_2 = \frac{q_f f_s}{1+q_f} (A_2-1)
\end{equation}
for S1 and S2, respectively. 

Table~\ref{tab:best-fit} provides the best-fit parameters of the four degenerate solutions, originating from the four-fold parallax degeneracy. These solutions have comparable values of $\chi^2$ and generally consistent values of xallarap parameters. In particular, the source binary is constrained to be nearly edge-on, which is a preferred configuration if one is to verify the 1L2S solution with radial velocity observations (see Section~\ref{sec:confirmation}). 

\begin{figure}
\includegraphics[width=1.\columnwidth]{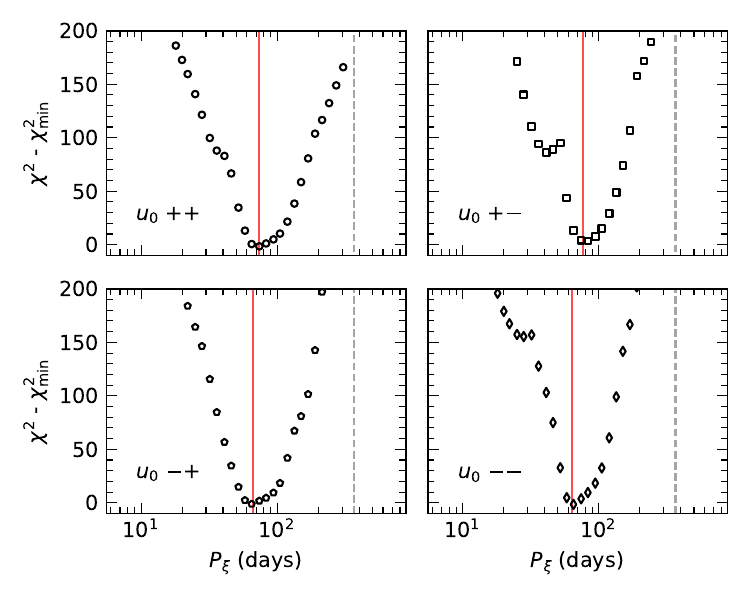}
\caption{
\revised{The change of $\Delta \chi^2$ versus $P_\xi$ of the four degenerate solutions. The $\chi^2_{\rm min}$ is the best-fitting $\chi^2$ of each solution, and the corresponding period is shown as the red solid line. The gray dashed line indicates the 1-year period. 
}
}
\label{fig:period_grid}
\end{figure}

\revised{The above xallarap solutions are identified through a free search in the whole parameter space. To ensure that we have not missed any degenerate solutions, a grid search in the orbital period of the source binary has been performed. In total 39 period values evenly spaced in logarithmic scale from 7 to 700\,days are selected, for each chosen period, we fix $P_\xi$ to the grid value and the parallax parameters to the globally best-fit values and search for the optimal parameter set that minimizes the model $\chi^2$ via MCMC. The results are shown in Figure~\ref{fig:period_grid} for all four solutions. Within the period range that was searched, we find no other solution that can fit the joint OGLE+Spitzer data equally well ($\Delta \chi^2\leq 100$). Therefore, the solutions presented in Table~\ref{tab:best-fit} are the only viable xallarap solutions of \eventname.}


\section{Physical Interpretation}
\label{sec:phy_interp}

\subsection{Source Properties}
\label{subsec:source_prop}

The angular Einstein radius can be determined by combining the xallarap amplitude, $\xi_{\rm E}$, and the semi-major axis of the primary source, $a_{\rm S}$
\begin{equation}
\label{eq:as_thetaE}
\theta_{\rm{E}} = \frac{a_{\rm{S}}}{D_{\rm{S}} \xi_{\rm{E}}} ,
\end{equation}
and $a_{\rm S}$ is related to the mass of the source binary via Kepler's third law
\begin{equation}
\label{eq:as1-def}
\frac{a_{\rm S}}{\rm AU} = \frac{q_\xi}{(1+q_\xi)^{2/3}} \left(\frac{M_{\rm S1}}{M_\odot}\right)^{1/3} \left(\frac{P_\xi}{\rm yr}\right)^{2/3}
\end{equation}
Here $M_{\rm S1}$ is the mass of the primary. The values of these masses are key in determining $\theta_{\rm E}$, as the xallarap amplitude and the source binary period are both measured in the xallarap modeling.

The mass of the source binary is estimated via the isochrone fitting. To perform this, we first derive the intrinsic colour and magnitude of both sources, following the general method of \citet{Yoo2004_OGLE2003BLG262}.
We first obtain the observed colours and magnitudes of both sources through a linear regression based on the 1L2S modeling of \eventname. This yields
\begin{equation} \label{eqn:S1-observed}
    \SonePhot
\end{equation}
for the primary source and
\begin{equation} \label{eqn:S2-observed}
    \StwoPhot
\end{equation}
for the secondary source. The above values are derived from the $(-,-)$ solution, i.e., the minimum $\chi^2$ solution, but the difference among the four degenerate solutions in these numbers is statistically negligible.

\begin{figure}
\includegraphics[width=1.\columnwidth]{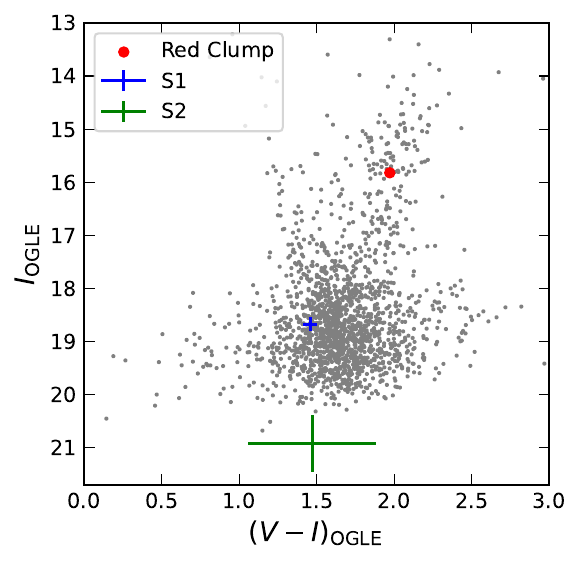}
\caption{
\revised{CMD for the $2\arcmin \times 2\arcmin$ square centred on \eventname. The best-fitting centroid of the red clump is shown as the red dot. The position of the S1 and S2 from solution $(-, -)$ are also marked on the figure. The blending is not shown, because it is not well constrained and generally consistent with zero. The relative positions of other degenerate solutions are the same as shown in Figure \ref{fig:CMD}.} 
}
\label{fig:RC-fitting}
\end{figure}

The centroid of the red clumps is determined to be 
\begin{equation}
\RCfit
\end{equation}
following the method of \cite{Nataf2013_Reddening}. \revised{We show the position of the red clumps centre as well as the S1, S2 from the $(-, -)$ solution in Figure \ref{fig:RC-fitting}. }At the location of \eventname, the intrinsic colour and dereddened magnitude of the red clump are 
\begin{equation}
\RCzero 
\end{equation}
adapted from \citet{Nataf2013_Reddening, Bensby2013_Chemical}, respectively. The reddening and extinction are therefore determined to be 
\begin{equation}
\Extinction. 
\end{equation}
Together with the observed colours and magnitudes of both sources (Equations~\ref{eqn:S1-observed} and \ref{eqn:S2-observed}) and by assuming the source is at $D_{\rm S}=8.2 \pm 1.4\,$kpc, we determine the intrinsic colour and absolute magnitude of both sources to be
\begin{equation}
\SonePhotZero
\end{equation}
and 
\begin{equation}
\StwoPhotZero
\end{equation}
respectively. For completeness, we have included in Table~\ref{tab:source-table} the measured and derived properties of the source binary for all four degenerate solutions.

\begin{figure}
\includegraphics[width=1.0\columnwidth]{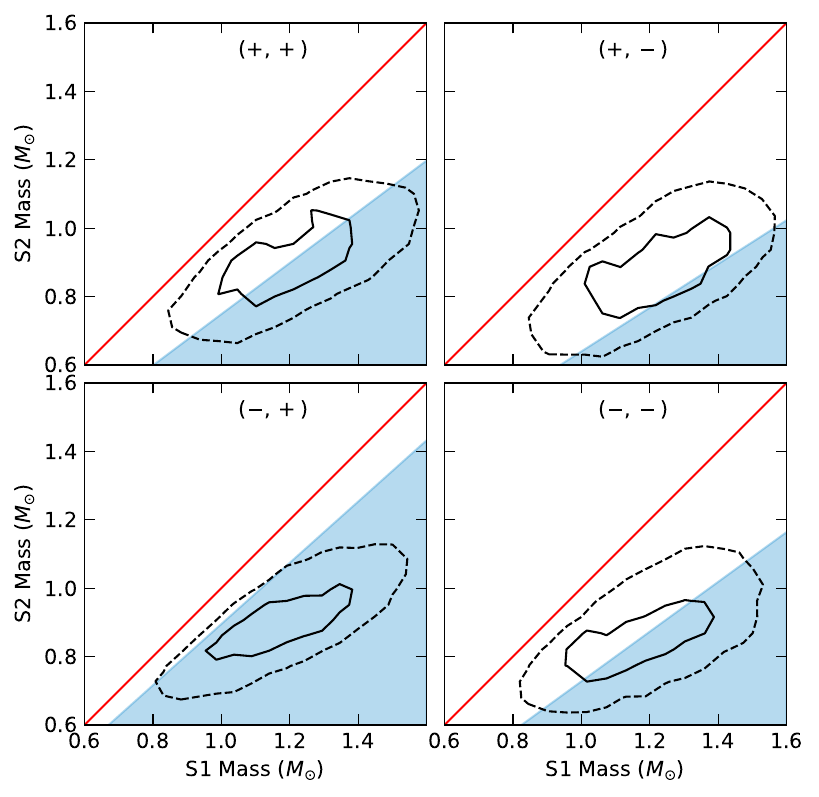}
\caption{
    Comparison between the binary mass from the isochrone fitting and the mass ratio $q$ from light curve fitting.
    The red line represents the $q=1$ line.
    The stellar masses determined by isochrone fitting are shown in black contours, with the solid and dashed contours showing the 1 and 2 sigma regions.
    The light blue shadow regions mark the mass ratio 90\% upper limit of each degenerate model.
}
\label{fig:q-iso}
\end{figure}

\begin{figure*}
\includegraphics[width=1.5\columnwidth]{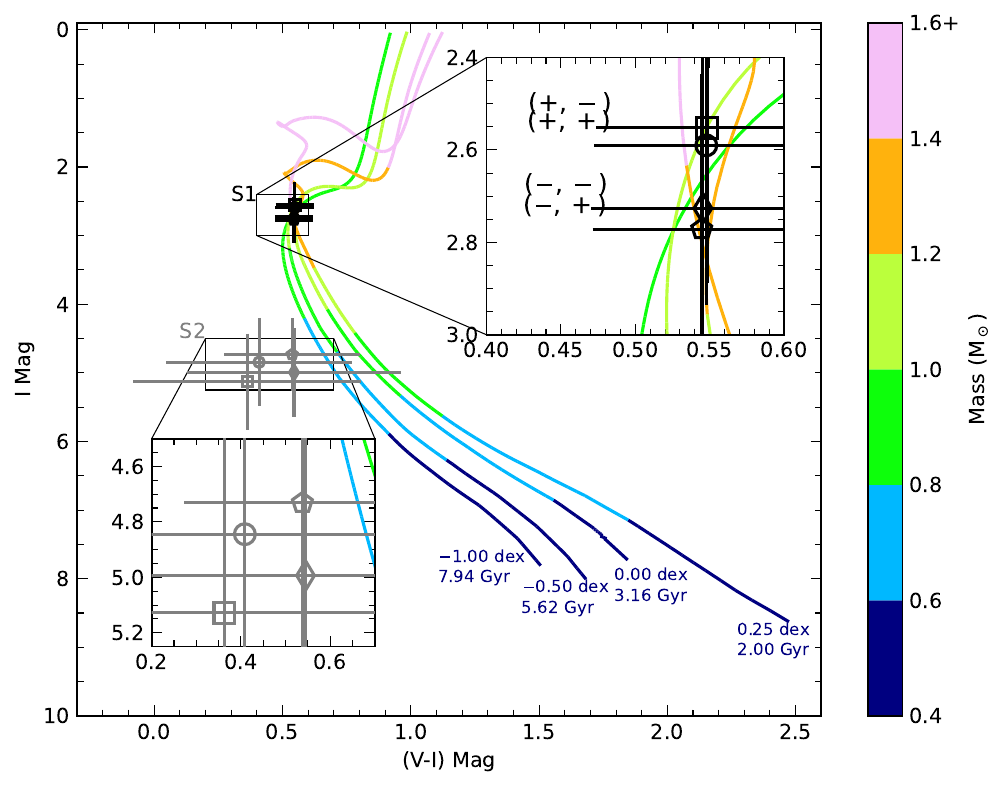}
\caption{
    The Colour-Magnitude diagram (CMD) of the four solutions with 1L2S xallarap + parallax model for \eventname as well as the isochrones.
    The four solutions are shown in different markers, i.e., the circle, square, pentagon, and diamond stand for solutions $(+,+)$, $(+,-)$, $(-, +)$, and $(-, -)$, respectively.
}
\label{fig:CMD}
\end{figure*}

The masses of the source stars are estimated based on their positions on the colour--magnitude diagram (CMD) with the \texttt{isochrones} package \citep{Morton2015_Isochrones}. This package adopts the MIST \citep{Dotter2016_MIST} isochrone and computes the log-likelihood of five model parameters, namely the equivalent evolutionary phase of both stars, the metallicity, age, and distance of the source system. Given the limited observational constraints, informative priors have been adopted. Specifically, the stellar metallicity is assumed to follow the observed distribution of the bulge microlensing stars from \citet{Bensby2017_Chemical}, and the distance is limited to the range 6.2--10.2\,kpc. The posterior distribution is sampled by \texttt{emcee} \citep{Foreman-Mackey2013_emcee}. The resulting constraints on the stellar masses are shown as the black contours in Figure~\ref{fig:q-iso}. 
For illustration purposes, we have shown four typical isochrones with different ages and metallicities that match the positions of the source stars, especially S1, in Figure~\ref{fig:CMD}. These isochrones are selected in the following way. First, four representative metallicity values, $(-1.0, -0.5, 0.0, 0.25)$, are selected based on the prior metallicity distribution function \citep{Bensby2017_Chemical}.
Then we determined the age of each isochrone so that it could match the colour and magnitude of S1 well.
This process yields an age range from $2.0$ Gyr to $7.9$ Gyr, and older ages always correspond to lower values of metallicity, which is also consistent with the observed age--metallicity trend of bulge stars \citep{Bensby2017_Chemical}.

The mass estimates from isochrone fitting also provide us with a way to verify the light curve modeling results. Figure~\ref{fig:q-iso} illustrates the masses of binary sources and the mass ratio constraints from light curve modeling (Table~\ref{tab:best-fit}) for all four degenerate solutions. The isochrone masses generally fall within the 1-$\sigma$ region of the mass ratio constraint, suggesting that these two sets of estimates are broadly consistent. Given a binary consisting of $\sim1.2$ and $\sim0.9 M_\odot$ stars, the flux ratio in \spitzer $3.6\mu$m band is around 0.4. This is in agreement with the flux ratio constraints for the $(-,+)$ and $(-,-)$ solutions and slightly disfavors (by $<3\sigma$) the $(+,+)$ and $(+,-)$ solutions. 

With the mass estimates of source binary stars, the semi-major axis of the source S1 is determined to be 0.4--0.5\,au, according to Kepler's third law. Together with the amplitude of the xallarap parameter, which is measured from the light curve modeling, we find the angular Einstein radius to be
\begin{equation}
    \theta_{\rm E} = 0.125 \left( \frac{a_{\rm S}}{0.2\,{\rm au}} \right) \left( \frac{D_{\rm S}}{8\,\rm kpc} \right)^{-1} \left( \frac{\xi_{\rm E}}{0.2}\right)^{-1} \rm mas.
\end{equation}
The exact values and uncertainties of these parameters are included in Table~\ref{tab:source-table} for all four degenerate solutions. \revised{For \eventname we use $D_{\rm S} = 8.2 \pm 1.4$ kpc, which is the distance and the radial extension of the bulge at the given galactic direction \citep{Nataf2013_Reddening}.}

\begin{table*}
\begin{threeparttable}
\caption{Physical properties of the source binary and lens object of \eventname.}
\renewcommand{\arraystretch}{1.4}
\begin{tabular}{lcccc}
\hline
\hline
           Parameters &              $(+,+)$ &              $(+,-)$ &              $(-,+)$ &              $(-,-)$ \\
\hline
$(V-I)_{\rm{S1, 0}}$ (mag) & $0.55 \pm 0.08$ & $0.55 \pm 0.07$ & $0.54 \pm 0.07$ & $0.55 \pm 0.08$ \\
$M_{I, \rm{S1}}$ (mag) & $2.6 \pm 0.3$ & $2.6 \pm 0.3$ & $2.8 \pm 0.3$ & $2.7 \pm 0.3$ \\
$(V-I)_{\rm{S2, 0}}$ (mag) & $0.4 \pm 0.4$ & $0.4 \pm 0.4$ & $0.5 \pm 0.3$ & $0.5 \pm 0.4$ \\
$M_{I, \rm{S2}}$ (mag) & $4.8 \pm 0.6$ & $5.1 \pm 0.7$ & $4.7 \pm 0.5$ & $5.0 \pm 0.6$ \\
$M_{\rm{S1}}$ ($M_{\sun}$) & $1.22 \pm 0.18$ & $1.23 \pm 0.18$ & $1.20 \pm 0.18$ & $1.20 \pm 0.18$ \\
$M_{\rm{S2}}$ ($M_{\sun}$) & $0.91 \pm 0.12$ & $0.88 \pm 0.13$ & $0.9 \pm 0.11$ & $0.87 \pm 0.12$ \\
$a_{\rm{S}}$ (AU) & $0.190 \pm 0.014$ & $0.189 \pm 0.014$ & $0.181 \pm 0.015$ & $0.177 \pm 0.017$ \\
$\theta_{\rm{E}}$ (mas) & $0.10^{+0.03}_{-0.02}$ & $0.091^{+0.022}_{-0.017}$ & $0.11^{+0.03}_{-0.02}$ & $0.11^{+0.03}_{-0.02}$ \\
$\mu_{\rm{rel, geo}}$ (mas/yr) & $0.84^{+0.21}_{-0.16}$ & $0.78^{+0.18}_{-0.15}$ & $0.93^{+0.25}_{-0.19}$ & $0.90^{+0.25}_{-0.19}$ \\
$M_{\rm L}$       &  $0.15^{+0.02}_{-0.02}$ &  $0.09^{+0.01}_{-0.01}$ &  $0.13^{+0.02}_{-0.02}$ &  $0.19^{+0.03}_{-0.03}$ \\
$D_{\rm L}$ (kpc) &  $7.68^{+0.06}_{-0.06}$ &  $7.54^{+0.07}_{-0.07}$ &  $7.46^{+0.10}_{-0.11}$ &  $7.72^{+0.06}_{-0.08}$ \\
\hline
\hline
\end{tabular}
\label{tab:source-table}
\end{threeparttable}
\end{table*}

\subsection{Mass and Distance of the Lens}
\label{subsec:lens_property}

Given the constraints on both $\theta_{\rm E}$ and $\pi_{\rm E}$, we now proceed to constrain the mass and distance of the lens object. The mass is given by
\begin{equation}
    M_{\rm L} = \frac{\theta_{\rm E}}{\kappa \pi_{\rm E}} \approx 0.15 \left( \frac{\theta_{\rm E}}{0.125\,\rm mas} \right) \left(\frac{\pi_{\rm E}}{0.1} \right)^{-1} M_\odot .
\end{equation}
Similarly, the lens--source relative parallax is given by
\begin{equation}
    \pi_{\rm rel} = \pi_{\rm E} \theta_{\rm E} = 0.0125 \left( \frac{\theta_{\rm E}}{0.125\,\rm mas} \right) \left(\frac{\pi_{\rm E}}{0.1} \right) \,{\rm mas} . 
\end{equation}
This value corresponds to a typical bulge lens. Adopting a source distance of $D_{\rm S}=8.2\,$kpc, we find the lens distance to be $D_{\rm L}\approx7.4$\,kpc. Therefore, the lens object of \eventname is a low-mass M-dwarf located in the bulge. The exact values and associated uncertainties of the derived mass and distance for all four degenerate solutions are also included in Table~\ref{tab:source-table}.

\begin{figure}
\includegraphics[width=1.\columnwidth]{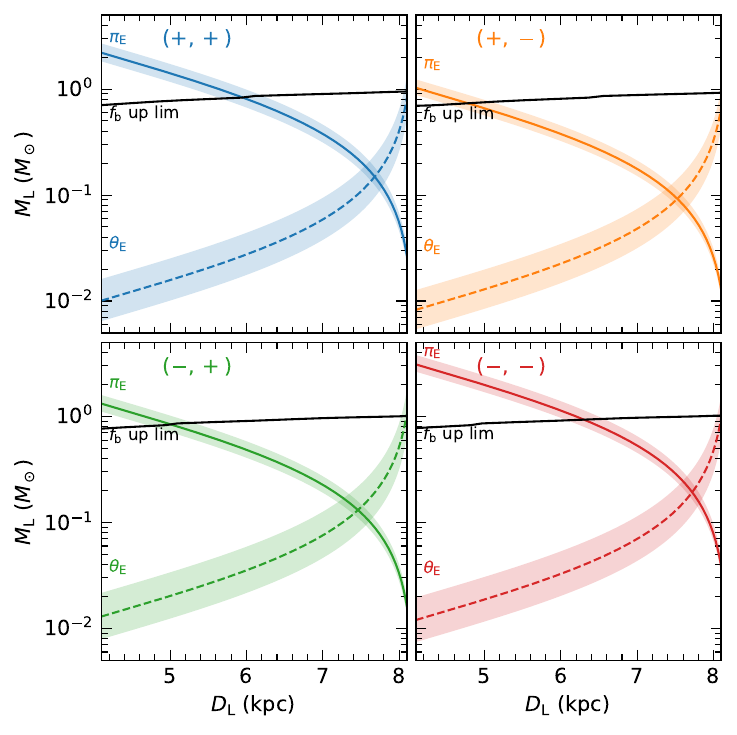}
\caption{
    The constraints on the lens mass and distance for \eventname, from $\pi_{\rm{E}}$,
    $\theta_{\rm{E}}$, and lens flux $f_{\rm{b}}$. The four degenerate solutions,
    $(+,+)$, $(+,-)$, $(-,+)$, and $(-,-)$, are shown in different panels.
    The black solid line in each panel is the upper limit from the lens flux.
    The coloured solid line and dashed line with a shaded region around them are the
    constraints from $\pi_{\rm{E}}$ and $\theta_{\rm{E}}$, respectively. 
}
\label{fig:mldl}
\end{figure}

Figure~\ref{fig:mldl} illustrates the constraints on the mass and distance of the lens object for the four degenerate solutions. In addition to the constraints from the angular Einstein radius and the microlensing parallax, we have also shown the upper limit from lens flux. This limit is derived by taking the 90\% upper limit on the $I$-band blending flux and assuming it is due entirely to the lens object. We have adopted the stellar mass--magnitude relations of \citet{Pecaut2012_Revised} and \citet{Pecaut2013_Intrinsic} \revised{and the extinction relation of \citet{Bennett2020_Keck}},
\begin{equation}
A_{\rm{L}} = (1 - \exp(-D_{\rm{L}} / \tau_{\text{dust}})) / (1 - \exp(-D_{\rm{S}} / \tau_{\text{dust}}))A_{\rm{S}}.
\end{equation}
\revised{Here $\tau_{\rm{dust}}=0.1 {\rm kpc}/\sin(|b|)$ is the scale length of the dust towards the galactic bulge}, with $b$ the Galactic latitude of the source system. For all four solutions, the lens mass and distance, derived from the parallax and angular Einstein radius measurements, are well within the allowed region by the lens flux upper limit.

\section{Discussion}
\label{sec:discuss}

\subsection{Impact of the Eccentricity}
\label{subsec:ecc}

In the 1L2S model with the xallarap effect, we have assumed a circular orbit for the source binary. However, observations have shown that binaries with eccentric orbits are fairly common \citep[e.g.,][]{Duchene2013_Stellar}. In the period range of 10--1000\,d, which is relevant for xallarap detection, about half of the binaries have eccentricities above $\sim$0.3 \citep{Raghavan2010_SURVEY}. Therefore, it is necessary to address the impact of eccentric orbits on the derived source and lens properties.

We adopt an eccentric orbit 1L2S model to fit the data. Together with the orbital eccentricity $e$, the argument of periapsis $\omega$ is also introduced to describe the eccentric motion. To avoid over-fit of the weak signal, we fix the orbital eccentricity to values in the range 0.1--0.8 and search the other parameters to minimize the model $\chi^2$. 
Compared to the circular model, the best $\chi^2$ values of the eccentric orbits are only marginally improved. At $e=0.3$, the improvement is $\Delta\chi^2\approx 5$. More eccentric ($e\gtrsim 0.4$) orbits have even worse $\chi^2$ values, and such orbits are disfavored \textit{a priori} at an orbital period of $\sim100\,$d.

With the introduction of eccentric orbits, model parameters that are relevant to the physical properties of the source and lens objects are not changed substantially. The changes in the flux ratios in different bands and in the mass ratio are all within the uncertainties of the corresponding parameters given in Table~\ref{tab:best-fit}. With the increasing eccentricity, the xallarap parameter and the orbital period of the source binary may both increase by up to 30--50\%, but because these two parameters have opposite effects on the angular Einstein radius, this latter parameter is not changed as much.
In the end, the mass and distance of the lens object are only changed by $\lesssim 20\%$, which is small compared to the difference between different degenerate solutions. It does not change the conclusion that the lens is a low-mass M-dwarf in the Bulge.

The inclination remains nearly unchanged with $e\leq 0.3$, but may reach up to $128^\circ$ at $e=0.8$. This has some implications for the radial velocity (RV) observations of the source binary, which will be discussed in the next section.

\subsection{Source Binary Confirmation} \label{sec:confirmation}

With a baseline magnitude of $I\approx18.5$ and \revised{a blending fraction that is low ($\lesssim 0.1$) and consistent with zero (Table~\ref{tab:best-fit})}, the source binary of \eventname is potentially accessible by spectroscopic observations from the ground. Such observations can not only confirm but also further refine the derived properties of the source binary (and thus the lens object).

\begin{figure*}
\includegraphics[width=1.8\columnwidth]{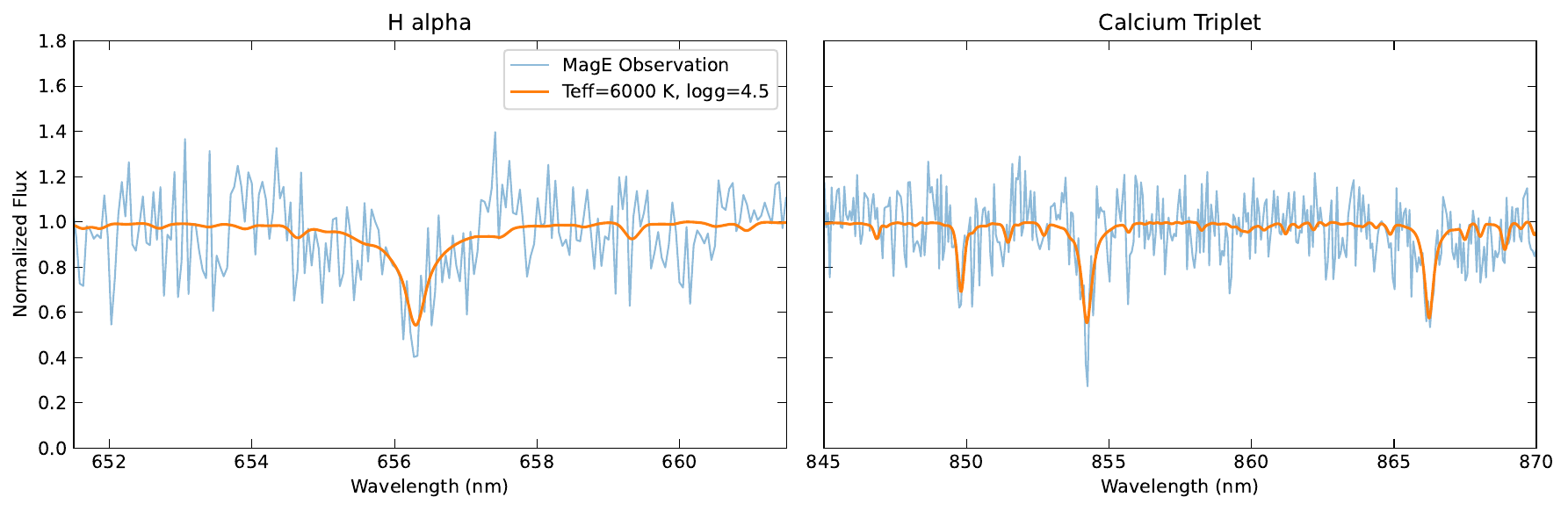}
\caption{
\revised{The observed MagE spectrum as well as the best-fitting stellar model. The MagE observation and best-fitting model are shown in blue and orange solid lines, respectively.}
}
\label{fig:spec}
\end{figure*}

We observed the source with the MagE spectrograph \citep{Marshall2008_MagE} on the 6.5 m Magellan Baade Telescope at Las Campanas Observatory on July 2, 2023. Three exposures, each with 20 min, were taken, and the resulting spectrum has a signal-to-noise ratio (S/N) per resolution of about five. Given the flux ratio of $\sim 0.15$, we expect to see only a single component in the spectra. We compared the observed spectrum with synthetic spectra generated with a stellar atmosphere grid MARCS \citep{Gustafsson2008_MARCS} and a radiative transfer code \texttt{turbospectrum} \citep{Plez2012_turbospectrum} integrated together by \texttt{iSpec} \citep{Blanco2019_iSpec}. The observed spectrum can be fitted with the spectra of late F-type subgiants and dwarfs reasonably well, while it disfavors the early F-type as well as the typical G-type stars. This is in agreement with the CMD fitting result. \revised{The observed spectrum, as well as the best-fitting stellar model near the H$\alpha$ and Ca triplet, are shown in Figure \ref{fig:spec}.}

Future spectroscopic observations are able to measure the RV variations of the source binary \citep[e.g.,][]{Ryu2024_Systematic}. Given the inferred parameters, the RV semi-amplitude is
\begin{equation}
K =28\  {\rm{km/s}} \frac{\sin i_\xi}{\sqrt{1 - e^2}} \left(\frac{M_{\rm{S2}}}{0.9 M_\odot}\right) \left(\frac{M_{\rm{S1}}+M_{\rm{S2}}}{2.1 M_\odot}\right)^{-2/3} \left(\frac{P_\xi}{75\ \rm{day}} \right)^{-1/3}.
\end{equation}
This is large enough for ground-based 10-m telescopes to achieve on a late F-type star. These RV observations will be able to directly measure the orbital eccentricity, which is currently not constrained in the model (Section \ref{subsec:ecc}). The combination of RV and xallarap can provide useful constraints on the angular Einstein radius, as has long been pointed out by \citet{Han1997_Einstein}.

As explained in Section~\ref{sec:1L2S}, the 1L2S model is favored mostly by the \spitzer data. It has been argued that the \spitzer data might suffer from systematics at some level \citep[e.g.,][]{Zhu2017_Galactic, Koshimoto:2020}, so the verification of the binary source via the RV method also provides an opportunity to further investigate the issues in the \spitzer data in this particular event.

\subsection{The Detectability of the Xallarap Effect}

As demonstrated by the event \eventname, the xallarap effect can provide additional information regarding the $\theta_{\rm{E}}$ parameter and thus has the capability to determine $\theta_{\rm{E}}$ by utilizing radial velocity measurement or isochrone fitting. Therefore, one wonders how often the xallarap effect appears and is detectable. In particular, under what conditions can the xallarap effect be distinguished from the annual parallax effect?

To answer the above question, we perform the following simplified simulations. We adopt the survey strategy and performance of \citet{Karolinski2020_Detecting} and set the $t_0$ is at the centre of an observing season. 
The values of $t_{\rm{E}}$ and $u_0$ were uniformly sampled in the ranges 10--150\,days and 0.1--1.0, respectively, which roughly follow the observed distribution of \citet{Mroz2019_Microlensing}. The angular Einstein radius is assumed to be 0.55 mas, corresponding to an M-dwarf lens located at 4\,kpc. The source binary consists of a primary star of mass $1\,M_\odot$ and a mass ratio of 0.3. This mass ratio corresponds to a luminosity ratio of $\lesssim 1\%$, sufficiently small that we can disregard the light of the secondary source in the light curve modeling. The orbital period of the source binary is sampled from a log-flat distribution between 0.5--10 times $t_{\rm{E}}$, and the orientation of the binary orbit and positions of the source stars are randomized.
For each of the simulated light curves, We have performed both parallax and xallarap modelings, and the xallarap signal is considered to be detected if the best-fit $\Delta \chi^2$ between parallax and xallarap models is $> 50$. 

\begin{figure}
\includegraphics[width=1.0\columnwidth]{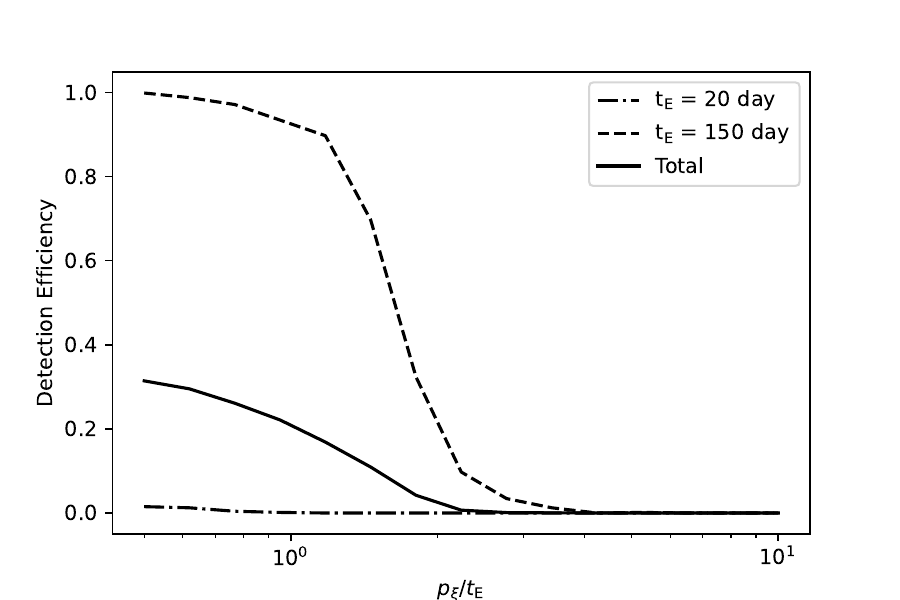}
\caption{
    The change of the xallarap effect detection efficiency with regard to $p_\xi/t_{\rm E}$ ratio when $u_0 = 0.1$. The solid line, dashed line, and dash-dotted line represent the total detection efficiency, as well as the detection efficiency when $t_{\rm E}$ is 150 or 20 days, respectively.
}
\label{fig:xlrp_DE}
\end{figure}

The detection efficiency of the xallarap effect is shown in Figure~\ref{fig:xlrp_DE} as a function of the binary orbital period. Source binaries with orbital periods $P\lesssim 2 t_{\rm E}$ are more likely to be detected, as their xallarap signals are difficult to be confused with annual parallax signals. Given the fixed sampling cadence, the detection efficiency therefore drops for shorter timescale events, resulting in an overall frequency of around 1\% once the binary period distribution is taken into account. This fraction increases for higher-cadence surveys and events with longer timescales. If we take roughly the same $t_{\rm E}$ distribution with the events in \citet{Poindexter2005_Systematic}, the detectability of the xallarap effect would be about 18\%, consistent with the result of \citet{Poindexter2005_Systematic} that about 23\% of the events with $t_{\rm E}\gtrsim 70$\,d are affected by xallarap effect. Because such long-timescale events are of particular interest in the search for dark lenses \citep[e.g.,][]{Lam:2020}, the impact of the xallarap effect and its usage in the lens mass determination should be evaluated seriously.

\section*{Acknowledgements}

\revised{We thank the reviewer for useful comments and suggestions.}
This work is supported by the National Natural Science Foundation of China (grant No.\ 12133005 \& 12173021) and the CASSACA grant CCJRF2105. \revised{R.~P.\ has been supported by the National Science Center, Poland, grant No. 2021/42/E/ST9/00038. For the purpose of Open Access, the author has applied a CC-BY public copyright license to any Author Accepted Manuscript (AAM) version arising from this submission.} We also acknowledge the Tsinghua Astrophysics High-Performance Computing platform for providing computational and data storage resources.

\section*{Data Availability}

The data underlying this article will be shared on reasonable request to the corresponding author.



\bibliographystyle{mnras}
\bibliography{main}  




\section*{Affiliations}

$^{1}$Department of Astronomy, Tsinghua University, Beijing 10084, China\\
$^{2}$Max-Planck-Institute for Astronomy, K\"onigstuhl 17, 69117 Heidelberg, Germany\\
$^{3}$Department of Astronomy, Ohio State University, 140 W. 18th Ave., Columbus, OH 43210, USA\\
$^{4}$Astronomical Observatory, University of Warsaw, Al. Ujazdowskie 4, 00-478 Warszawa, Poland\\
$^{5}$Department of Earth and Space Science, Graduate School of Science, Osaka University, Toyonaka, Osaka 560-0043, Japan\\
$^{6}$Department of Particle Physics and Astrophysics, Weizmann Institute of Science, Rehovot 7610001, Israel.\\
$^{7}$IPAC, Mail Code 100-22, Caltech, 1200 E. California Blvd., Pasadena, CA 91125, USA\\
$^{8}$Center for Astrophysics, Harvard \& Smithsonian, 60 Garden Street, Cambridge, MA 02138, USA \\
$^{9}$Jet Propulsion Laboratory, California Institute of Technology, 4800 Oak Grove Drive, Pasadena, CA 91109, USA\\
$^{10}$Department of Physics and Astronomy, Texas Tech University, 1200 Memorial Circle, Lubbock, TX 79409, USA\\
$^{11}$Department of Physics and Astronomy, Michigan State University, 567 Wilson Road, East Lansing, MI 48824, USA\\
$^{12}$Department of Physics, University of Warwick, Gibbet Hill Road, Coventry, CV4 7AL, UK\\
$^{13}$Institute for Space-Earth Environmental Research, Nagoya University, Nagoya 464-8601, Japan\\
$^{14}$Code 667, NASA Goddard Space Flight Center, Greenbelt, MD 20771, USA\\
$^{15}$Department of Astronomy, University of Maryland, College Park, MD 20742, USA\\
$^{16}$Institute of Natural and Mathematical Sciences, Massey University, Auckland 0745, New Zealand\\
$^{17}$Department of Earth and Planetary Science, Graduate School of Science, The University of Tokyo, 7-3-1 Hongo, Bunkyo-ku, Tokyo 113-0033, Japan\\
$^{18}$Instituto de Astrof\'isica de Canarias, V\'ia L\'actea s/n, E-38205 La Laguna, Tenerife, Spain\\
$^{19}$Institute of Astronomy, Graduate School of Science, The University of Tokyo, 2-21-1 Osawa, Mitaka, Tokyo 181-0015, Japan\\
$^{20}$Department of Physics, The Catholic University of America, Washington, DC 20064, USA\\
$^{21}$Institute of Space and Astronautical Science, Japan Aerospace Exploration Agency, 3-1-1 Yoshinodai, Chuo, Sagamihara, Kanagawa 252-5210, Japan\\
$^{22}$Sorbonne Universit\'e, CNRS, UMR 7095, Institut d’Astrophysique de Paris, 98 bis bd Arago, 75014 Paris, France\\
$^{23}$Department of Physics, University of Auckland, Private Bag 92019, Auckland, New Zealand\\
$^{24}$University of Canterbury Mt. John Observatory, P.O. Box 56, Lake Tekapo 8770, New Zealand\\

\appendix

\section{Binary Lens Interpretation}
\label{app:a1_2L1S}

An additional companion to the lens object may also lead to distortions in the microlensing light curve \citep{Mao1991_Gravitational, Gould1992_Discovering}. As long as the source star stays relatively away from the caustics of the binary lens, such distortions are not prominent features and may resemble those produced by the xallarap effect \citep[e.g.,][]{Rota2021_MOA-2006-BLG-074, Yang:2024}. We therefore carry out 2L1S modeling including the lens orbital motion to make sure the detected signal is indeed caused by the xallarap effect.

\begin{figure*}
\includegraphics[width=1.9\columnwidth]{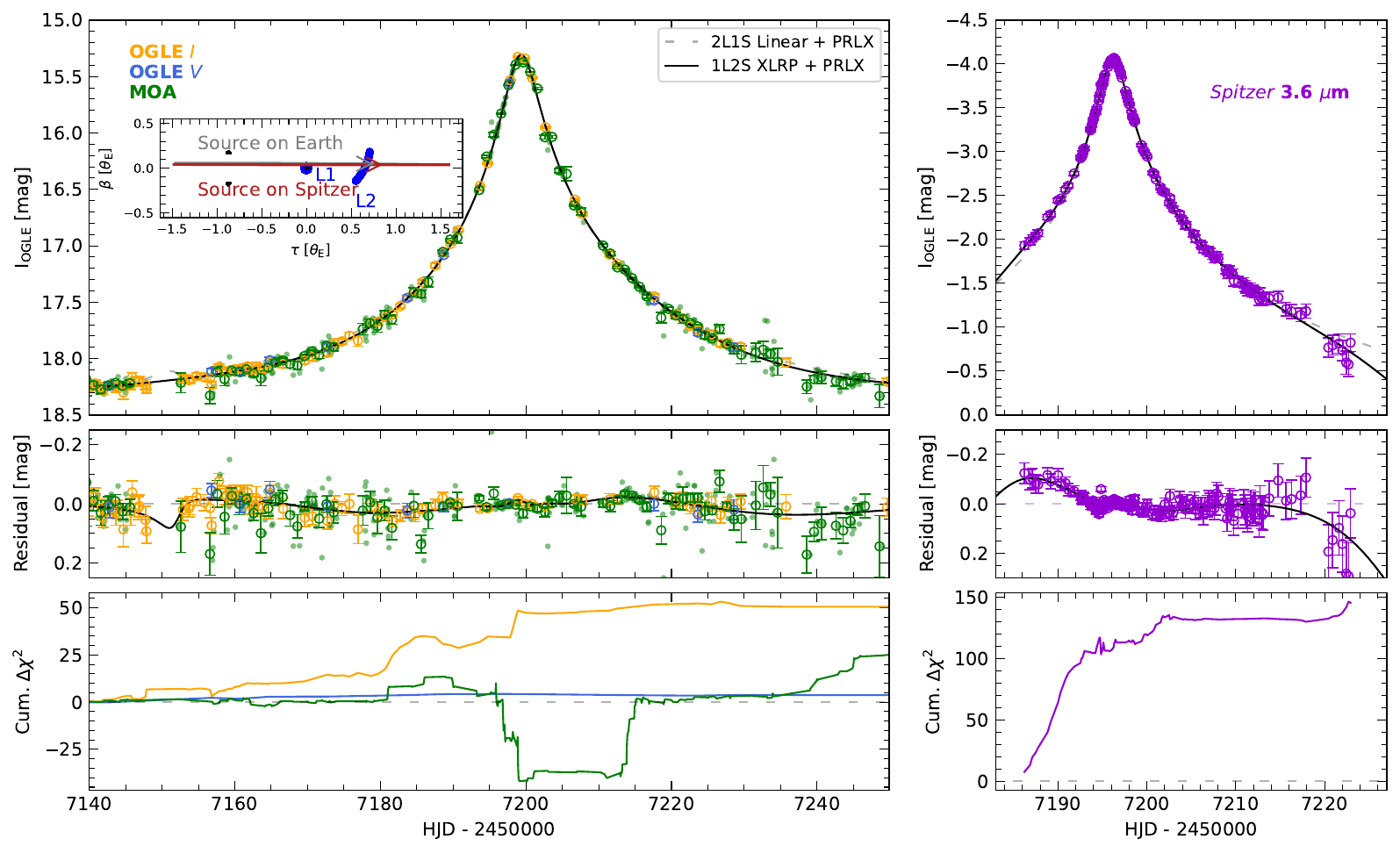}
\caption{
    The comparison between the 2L1S model and the 1L2S xallarap + parallax model. The figure description is similar to that of Figure~\ref{fig:lc-all}. In the figure, \revised{we present both the original MOA data without error bars and the MOA data binned into daily cadence, which are denoted as green dots and circles with error bars, respectively.}.
    The inset in the upper panel shows the trajectory of the source as well as the caustic curve of the binary lens and the position of the lens.
}
\label{fig:lc_binl_test}
\end{figure*}

In this 2L1S modeling, we \revised{have included the MOA data as well based on the arguments in Section~\ref{sec:obs}.} 
To describe the 2L1S model, \revised{six} new parameters are introduced \revised{in addition to the 1L1S parameters, including four standard parameters ($\rho$, $\alpha$, $s$, and $q$) and two that describe the linearized lens orbital motion (${\rm d} \alpha / {\rm d} t$ and ${\rm d} s / {\rm d} t$)}. 
Here, $\rho$ represents the angular size of the source normalized to the angular Einstein radius, $\alpha$ is the angle between the source trajectory and the binary axis, and $s$ and $q$ are the projected separation and mass ratio between the two lens components, respectively. \revised{The orbital motion of the lens system is included in order to have a fair comparison with the binary source model with the xallarap effect.}
Following a thorough grid search and refined MCMC sampling of the posterior distribution, we only identify one binary lens solution that matches the observed data the best. This best-fit 2L1S model is shown in Figure~\ref{fig:lc_binl_test}, and the parameter values are given in Table~\ref{tab:2L1S} for the purpose of completeness. As shown in Figure~\ref{fig:lc_binl_test}, the best-fit 2L1S model is not able to explain the deviations in the ground-based data fully, and it is worse than the 1L2S xallarap model by $\Delta \chi^2$ of $\sim 230$. \revised{Here the 1L2S xallarap model is the same as given in Table~\ref{tab:best-fit}, and we have introduced a flux ratio in the MOA band to fit the MOA data. Given the relatively large $\Delta \chi^2$ value, }
we therefore exclude the binary lens model as a plausible solution for \eventname.

\begin{table}
\centering
\begin{threeparttable}
\caption{\revised{Parameters of the best-fit 2L1S solution for \eventname}.}
\renewcommand{\arraystretch}{1.4}
\begin{tabular}{lc}
\hline
Parameters               & $(+,+)$                \\ \hline
$t_{0}$ (HJD$^\prime$)   & $7199.6348 \pm 0.0087$  \\
$u_{0}$                  & $0.0547 \pm 0.0010$   \\
$t_\mathrm{E}$ (days)    & $40.5 \pm 0.5$        \\
$\pi_\mathrm{E, N}$      & $-0.0030 \pm 0.0003$        \\
$\pi_\mathrm{E, E}$      & $0.0797 \pm 0.0011$        \\
$\rho$                   & $0.008 \pm 0.003$      \\
$\alpha$ ($\deg$)        & $7 \pm 52$       \\
$s$      & $0.641 \pm 0.010$       \\
$q$       & $0.0058 \pm 0.0003$       \\
${\rm d} \alpha / {\rm d} t $ (deg / yr) & $-88 \pm 8$       \\
${\rm d} s / {\rm d} t $ (yr$^{-1}$) & $0.44 \pm 0.08$       \\
Blend Fractioni          & $0.089 \pm 0.016$      \\
$\chi^2$/dof             & 3095.11/2923            \\ \hline
\end{tabular}
\begin{tablenotes}
    \item{
    NOTE. HJD$^\prime$=HJD-2450000.
    }
\end{tablenotes}
\label{tab:2L1S}
\end{threeparttable}
\end{table}

\section{Static Binary Source Interpretation}
\label{app:a2_static_1L2S}

\revised{\revisedthree{In Table \ref{tab:1L2S_static}, we present the best-fit binary source solution without orbital motion (i.e., 1L2S static) here for completeness.} In the 1L2S static model, the position of the secondary source is given by $t_{0, 2}$ and $u_{0, 2}$, which are the time of closest approach and the impact parameter of the secondary source, respectively \citep[e.g.,][]{Hwang2013_Interpretation}. Similar to the 1L2S xallarap model, the flux ratios between the two sources in individual bandpasses are also required.}

\revised{The comparison between the best-fit 1L2S static solution and the best 1L2S xallarap solution is shown in Figure \ref{fig:lc_bins_test}, with the latter one the same as in Appendix \ref{app:a1_2L1S}. Similar to the origin of the four-fold parallax degeneracy, there exist multiple degenerate solutions for the 1L2S xallarap explanation, and here we select the one with the smallest model $\chi^2$. 
As shown in Figure~\ref{fig:lc_bins_test}, even this best-fit 1L2S static solution is worse than the best 1L2S xallarap solution by $\Delta \chi^2 > 230$, and it fails to explain the general behaviour of the light curve between $7160 < {\rm HJD}\arcmin < 7195$. Therefore, the 1L2S static model is excluded.}

\begin{table}
\centering
\begin{threeparttable}
\caption{\revisedthree{Parameters of best-fit 1L2S static solution for \eventname}.}
\renewcommand{\arraystretch}{1.4}
\begin{tabular}{lc}
\hline
Parameters                              & $(+, -)$                \\ \hline
$t_{0,1}$ (HJD$^\prime$)                & $7199.376 \pm 0.006$ \\
$u_{0,1}$                               & $0.0666 \pm 0.0007$  \\
$t_\mathrm{E}$ (days)                   & $31.8 \pm 0.3$       \\
$\pi_\mathrm{E,N}$                      & $-0.0968 \pm 0.0010$ \\
$\pi_\mathrm{E,E}$                      & $0.1129 \pm 0.0010$  \\
$t_{0,2}$ (HJD$^\prime$)                & $7207.6 \pm 0.4$     \\
$u_{0,2}$                               & $0.230 \pm 0.013$    \\
$q_\mathrm{f,OGLE}$                     & $0.137 \pm 0.011$    \\
$q_\mathrm{f,OGLE,V}$                   & $0.106 \pm 0.023$    \\
$q_\mathrm{f,MOA}$                      & $0.137 \pm 0.011$    \\
$q_\mathrm{f,Spitzer}$                  & $0.037 \pm 0.007$    \\
Blend Fraction                          & $-0.265 \pm 0.017$   \\
$\chi^2$/dof                            & 3104.71/2921         \\
\hline
\end{tabular}
\begin{tablenotes}
    \item{
    NOTE. HJD$^\prime$=HJD-2450000.
    }
\end{tablenotes}
\label{tab:1L2S_static}
\end{threeparttable}
\end{table}

\begin{figure*}
\includegraphics[width=1.9\columnwidth]{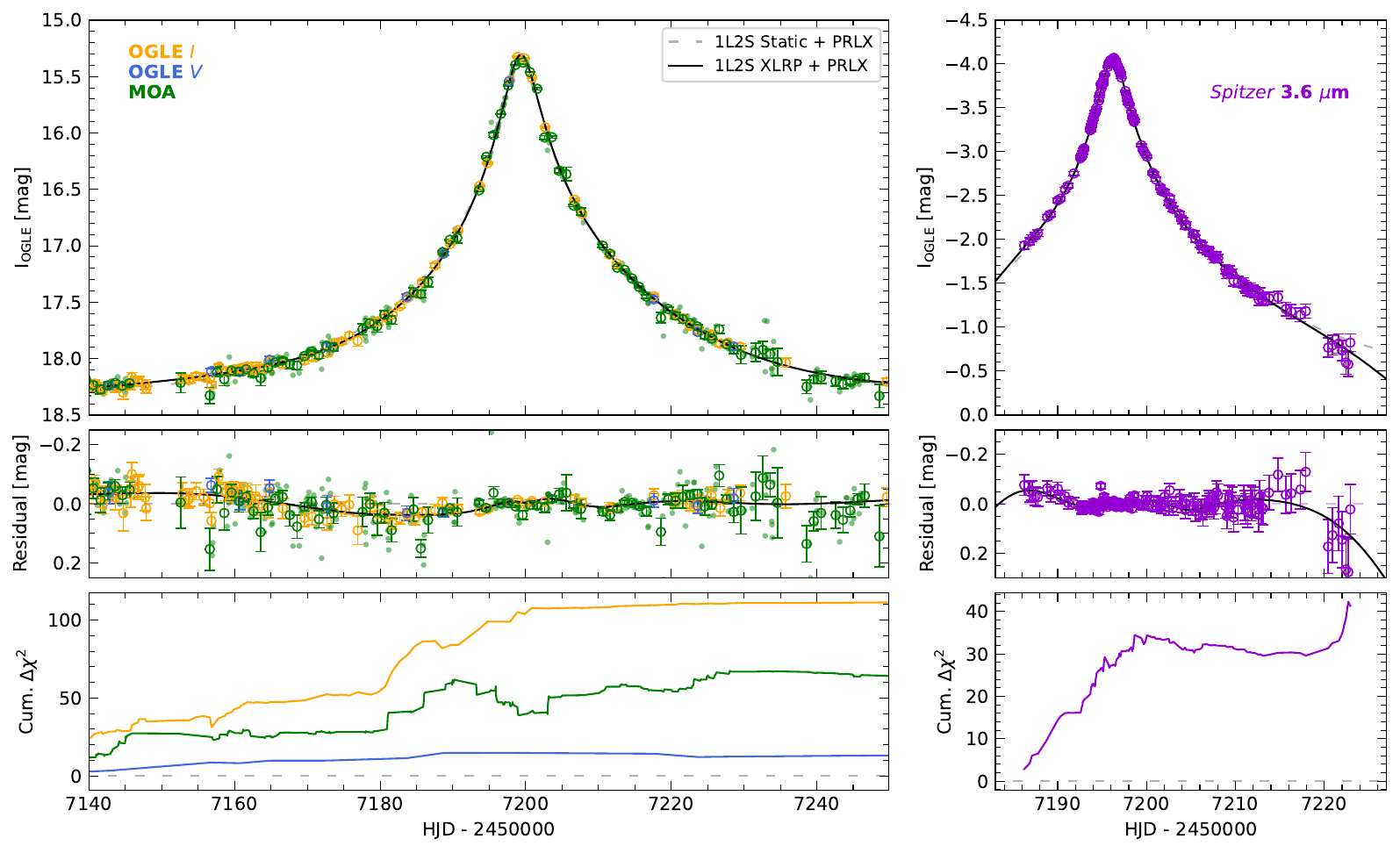}
\caption{
    \revised{The comparison between the 1L2S static and the 1L2S xallarap model. The figure description is similar to that of Figure \ref{fig:lc_binl_test}.}
}
\label{fig:lc_bins_test}
\end{figure*}


\bsp	
\label{lastpage}
\end{document}